\def\simgr{\mathbin{\;\raise1pt\hbox{$>$}\kern-8pt\lower3pt\hbox{$\sim$}\;}}
\def\simlr{\mathbin{\;\raise1pt\hbox{$<$}\kern-8pt\lower3pt\hbox{$\sim$}\;}}
\documentclass{aa}
\usepackage{epsfig}
\usepackage{amssymb}

\begin{document}
\thesaurus{08.18.1, 08.05.3, 08.01.1, 08.14.1, 08.05.1} 
\title{Stellar Evolution with Rotation V: Changes in all the Outputs of
Massive Star Models}

\author{Georges Meynet  \and Andr\'e Maeder}
   
   \offprints{G. Meynet}

   \institute{Geneva Observatory,
              University of Geneva, 
              CH--1290 Sauverny,
	      Switzerland}
   \titlerunning{Models for rotating massive stars}
   \maketitle
   
   \markboth{G. Meynet, A. Maeder: Stellar Evolution with Rotation}
{G. Meynet, A. Maeder: Stellar Evolution with Rotation}

\begin{abstract}

Grids of  models for rotating stars  are constructed
in the range of 9 to 120 M$_{\odot}$
at solar metallicity. The following effects of rotation are included:
shellular rotation, new structure equations for non--conservative case,
surface distorsions, increase of mass loss with rotation, meridional
circulation and interaction with horizontal turbulence, shear instability
and coupling with thermal effects, advection and diffusion of angular
momentum treated in the non--stationary regime, transport and diffusion
of the chemical elements. 

Globally we find that for massive
stars the effects of rotation have an importance comparable to those
of mass loss. Due to meridional circulation the internal rotation law $\Omega(r)$
rapidly converges, in 1--2 \% of the MS lifetime, towards a near equilibrium
profile which then slowly evolves during the MS phase. The circulation
shows two main cells. In the deep interior, circulation rises along
the polar axis and goes down at the equator, while  due to the
Gratton--\"{O}pik term it is the inverse in  outer layers. This
external inverse circulation grows in depth as evolution proceeds. We
emphasize that a stationary approximation and
a diffusive treatment of meridional circulation
would be unappropriate. After the MS phase,
the effects of core contraction and envelope expansion dominate
the evolution of the angular momentum. 

The surface velocities 
decrease very much during the MS evolution of the most massive stars, 
due to
their high mass loss, which also removes a lot of angular momentum. This produces
some convergence of the velocities, but not necessarily towards the
break--up velocities. However, stars with masses below $\sim$12 M$_{\odot}$
with initially high rotation may easily reach the break--up
velocities near the end of the MS phase, which may explain the 
occurence of Be--stars. Some other interesting properties of the 
rotational velocities are pointed out.

For an average rotation, the tracks in the HR diagram are modified
like a moderate overshoot would do. In general, an average rotation
may increase the MS lifetime up to about 30 \%; for the 
helium--burning  phase 
the effects are smaller and amount to at most 10 \%. 
From plots of the isochrones, we find that rotation may increase
the age estimate by about 25 \% in general.
However, for stars with 
M $\simgr$40 M$_{\odot}$ and  fast rotation, a bluewards
``homogeneous--like'' track, with important He-- and N--enrichments,
may occur drastically affecting the age estimates for the youngest 
clusters. Rotation also introduces a large scatter in
the mass--luminosity relation: at the same $\log g_{\rm eff}$ and
$\log T_{\rm eff}$, differences of masses
by 30 \% may easily occur, thus explaining what still remains
of the alleged mass discre\-pancy.

Rotation also brings significant surface He-- and N--enhancements, 
they are higher for higher masses and rotation. While it is not
difficult to explain very fast rotators with He-- and N--excesses, 
the present models also well account for the many OB stars
exhibiting surface enrichments and moderate or low rotation,
(cf. Herrero et al. \cite{He92}, \cite{He20}). These stars likely
result from initially fast rotators, which experienced mixing
and lost a lot of angular momentum due to enhanced mass loss.
The comparison of the N--excesses for B-- and A--type supergiants 
supports the conclusion by Venn (\cite{ve95a}, \cite{Ve99}), that these
enrichments 
mostly result from mixing during the MS phase, which is also 
in agreement with the results of Lyubimkov (\cite{Lyu96}).

\end{abstract}


\section{Introduction}

Due to the many well known successes of the  standard
theory of stellar evolution, rotation has generally been considered as 
a secondary effect. This was justified in many cases.
However, since some years a number of very signi\-ficant discrepancies
between model predictions and observations have been found for massive
stars, and also for red giants of lower masses. 
These difficulties have been listed and examined
(cf. Maeder \cite{Mae95a}). Recently, large excesses of
[N/H] have been found in A--type supergiants, particularly in the
SMC (Venn et al. \cite{Ve98}; Venn \cite{Ve99}), where the excesses may reach up to
an order of magnitude  with respect to the 
average [N/H] local value in the SMC. These excesses, not
predicted by current models, are the signatures of
 mixing effects over the entire star. This extensive  mixing
changes the size of the reservoir of nuclear fuel available for
evolution, and thus the lifetimes, the tracks and all the
model outputs
(cf. also Langer 1992; Langer 1997; Heger et al. 2000). Also the chemical abundances, the yields and
the final stages may be modified. 
Rotation appears as a natural driver for this mixing or, at least,
as one of the first mechanisms whose consequences for mixing has to be explored.
This is especially true for massive stars which are known to be fast rotators
(see e.g. Penny 1996; Howarth et al. 1997). 

The physical effects  of stellar rotation are numerous. The basic equations
of stellar structure need to be mo\-dified. The meridional circulation
and its interaction with the horizontal turbulence, the diffusion effects
produced by shear turbulence, the inhibition of the transport me\-chanisms
by the $\mu$--gradients,
the transport of the angular momentum and of the chemical elements, the loss
of mass and angular momentum at the surface, the
enhancement of mass loss by rotation, etc\dots\  are some of the effects
to be included in realistic models. In addition to the model physics,
rotation also brings a number of new numerical problems in the stellar
code, such as the 4$^{th}$ order equation for the transport of 
the angular momentum, which has to be coupled to the equations of stellar structure.

In this work, we apply the investments in the treatment of
the physical and numerical effects of rotation made 
in the previous works and we
explore their  consequences for the outputs of stellar models.
In Sect. 2, we describe the various effects considered. 
The non--rotating stellar models are briefly discussed in Sect. 3.
The evolution
of the internal rotation during the evolution is examined in Sect. 4.
Sect. 5 discusses the effects of rotation on the evolutionary tracks in the HR diagram, 
on the lifetimes and isochrones. The evolution of the rotational velocities
at the stellar surface
is discussed in Sect. 6 and the effects on the surface abundances are
analysed in Sect. 7.

\section{Physical ingredients of the models}

Let us briefly summarize here the basic physical ingredients of 
the numerical models of rotating stars we  are constructing here.
 
\subsection{Shellular rotation}
The differential rotation which results from the evolution and transport of the
angular momentum as described by Eq.~(3) below,
makes the stellar
interior highly turbulent. The turbulence is  very anisotropic, with
a much  stronger geostrophic--like transport
in the horizontal direction than in the vertical one (Zahn
\cite{Za92}),
where stabilisation is favoured by the stable temperature gradient.
This strong horizontal transport is characterized by a 
diffusion coefficient $D_{\rm{h}}$, which is quite large as
will be shown below.
The horizontal turbulent coupling  favours an essentially 
constant angular velocity $\Omega$   on the isobars. 
This rotation law, constant on shells,  applies to fast as well as to slow
rotators. As an approximation, it is often represented by a law of 
the form $\Omega = \Omega (r)$ (Zahn \cite{Za92}; see also Endal and Sofia \cite{ES76}).
 
\subsection{Hydrostatic effects}
In a rotating star, the equations of stellar structure
 need to be modified  (Kippenhahn and Thomas \cite{KippTh70}).
The usual spherical coordinates must be replaced by new coordinates 
characterizing the equipotentials.
The classical method applies when  the effective
gravity can be derived from a potential
$\Psi = \Phi - \frac{1}{2} \Omega^2 r^2 \sin^2 \theta$,
 i.e. when the  problem is conservative. There, $\Phi$ is the
gravitational potential which in the Roche approximation is $\Phi=-\frac{G M_r}{r}$.
If  the rotation law is shellular, the problem is non--conservative.
Most existing models of rotating stars apply, rather
inconsistently, the classical scheme by Kippenhahn and Thomas.
However, as shown by Meynet and Maeder (\cite{MM97}),
the equations of stellar structure can still be 
written consistently, in term of a coordinate referring to 
the mass inside the isobaric surfaces. Thus, the problem
of the stellar structure
of a differentially rotating star with an angular velocity
$\Omega = \Omega(r)$ can be kept one--dimensional.

\subsection{Surface conditions}
 The distribution of temperature at
 the surface of a rota\-ting star is described by
the von Zeipel theorem (\cite{vZ24}).
Usually, this theorem applies to the conservative case and states that 
the local radiative flux
 $\vec{F}$ is proportional to the local effective gravity
$\vec{g_{\rm eff}}$, which  is the sum of the 
gravity and centrifugal force,

\begin{equation}
\vec{F} = - \frac{L(P)}{4\pi GM_\star(P)} \, \vec{g_{\rm eff}} \; ,
\end{equation}

\noindent with $M_{\star}(P) = M(1 - \frac{\Omega^2}{2\pi G\bar{\rho}})$;
$L(P)$ is  the luminosity on an isobar and $\bar{\rho}$ the mean
internal density. The local $T_{\rm eff}$ on the surface of a rotating
star  varies like
$T_{\rm eff}(\vartheta) \sim g_{\rm eff}(\vartheta)^{\frac{1}{4}}$. 
We define the average stellar $T_{\rm eff}$  by
 $T_{\rm eff}^4= L/(\sigma S(\Omega))$, 
where $\sigma$ is Stefan's constant and $S(\Omega)$ the total actual
stellar surface. Of course, for different orientation angles
$i$, the emergent luminosity, colours and spectrum will be 
different (Maeder and Peytremann \cite{MP70}).
In the case of non--conservative rotation law, the corrections
to the von Zeipel theorem depend on the opacity law and 
on the degree of differential rotation, but they are  small,
i.e. $\leq 1 \%$ in current cases of shellular rotation
(Kippenhahn \cite{Kipp77}; Maeder \cite{Mae99}).
Wether or not the star is 
close to the Eddington limit, the 
von Zeipel theorem keeps the same form as the one given by Maeder
(1999, see also Maeder \& Meynet \cite{Mae00}). In this last work, it is in 
particular shown that the expression for the Eddington factor 
in a rotating star needs to be consistently written and that it contains a term depending on rotation.

\subsection{Changes of the mass loss rates $\dot{M}$ with  rotation}
 Observationally, a growth of the mass flux
of OB stars with rotation, i.e.\ by 2--3 powers of 10, was found
by Vardya (\cite{Var85}), while   Nieuwenhuijzen and de Jager
(\cite{Nieu88}) concluded that the $\dot M$--rates 
seem to increase only slightly
with rotation for O-- and B--type stars. 
On the theoretical side, 
Friend and Abbott (1986) find an increase of the $\dot{M}$--rates
which can be fitted by the relation (Langer \cite{La98})

\begin{equation}
\dot{M}(v)\; = \;\dot{M} (v = 0) 
\left(\frac{1}{1-\frac{v}
{v_{\rm crit}}}\right)^{\xi}
\end{equation} 

\noindent with $\xi \simeq 0.5$ and $v_{\rm crit}$ the equatorial
velocity at break--up. This 
expression is used in most evolutionary models and this is also
what is done in the present work. The critical rotation velocity of a star is often written
as $v^2_{\rm{crit}} = \frac{GM}{R_{\rm eb}} (1-\Gamma)$, where
$R_{\rm eb}$ is the equatorial radius at break--up velocity and
$\Gamma = L/L_{\rm{Edd}}$ is the ratio of the 
stellar luminosity to the Eddington luminosity (cf. Langer 1997, 1998).
Glatzel (1998) has shown that when
the effect of gravity darkening is taken into account, 
the above expression for $v_{\rm{crit}}$ does not apply.
Glatzel (1998) gives
$v^2_{\rm{crit}} = \frac{GM}{R_{\rm eb}}$, which we adopt here.
The problem is now being further examined  by Maeder \& Meynet (2000a),
who critically discuss the Eddington factors,
their dependence on rotation, the expression of the critical velocity, 
the dependence of the mass loss rates on rotation. These various new results
will be applied in subsequent works, particularly for the study of the effects
of rotation on the formation of W--R stars.
Further improvements, based
either on the observations or on the theory, to
account for the  anisotropic winds which selectively remove
the angular momentum need also to be performed. 

The above
expression gives the change of the mass loss rates due to rotation.
As reference mass loss rates in the case of no rotation, 
we use the recent data by Lamers and Cassinelli (\cite{LaCa96});
for the domain not covered by these authors
we use the results by de Jager et al. (\cite{Ja88}).
During the Wolf--Rayet phase we use
the mass loss rates proposed by Nugis et al (1998) for the
WNL stars (mean, clumping--corrected rates from radio data $\dot M({\rm WNL}) = 3\  10^{-5}$ M$_\odot$ y$^{-1}$).
For the WNE and WC stars we use the prescription devised by Langer (1989), 
modified according to Schmutz (1997) for taking into account the
clumping effects in Wolf-Rayet stellar winds ($\dot M = 2.4\cdot  10^{-8} (M/{\rm M}_\odot)^{2.5}$ M$_\odot$ y$^{-1}$).
These mass loss rates  
are smaller by a factor 2--3 than the mass loss rates used in our previous stellar grids
(Schaller et al. 1992; Meynet et al. 1994). 

\subsection{Transport of the angular momentum}
For shellular rotation, the equation of transport of angular
momentum in the vertical direction is in lagrangian coordinates
(cf. Zahn \cite{Za92}; Maeder and Zahn \cite{MZ98})

\begin{eqnarray}
\lefteqn{\rho \frac{d}{d t}
\left( r^2 \Omega\right)_{M_r} = } \nonumber \\[2mm]
&& \frac{1}{5 r^2}  \frac{\partial}{\partial r}
\left(\rho r^4 \Omega U(r) \right)
  + \frac{1}{r^2} \frac{\partial}{\partial r}
\left(\rho D r^4 \frac{\partial \Omega}{\partial r} \right) .
\end{eqnarray} 

\noindent $\Omega(r)$ is the mean angular velocity at level $r$.
The vertical component $u(r,\theta)$ of the velocity of the meridional
circulation at a distance $r$ to the center and at a colatitude $\theta$ can
be written

\begin{eqnarray}
u(r,\theta)=U(r)P_2(\cos \theta),
\end{eqnarray} 

\noindent where $P_2(\cos \theta)$ is the second Legendre polynomial. Only the radial term
$U(r)$ appears in Eq.~(3).
The quantity $D$ is  the  total diffusion coefficient representing
the various instabilities considered  and which transport
the angular momentum, namely  convection, 
semiconvection and  shear turbulence. 
As a matter of fact, a
very large diffusion coefficient as in convective regions
implies a rotation law which is not far from solid body
rotation. In this work, we take  
$D = D_{\rm{shear}}$ in
radiative zones, since as extra--convective mixing
we  consider shear mixing and meridional circulation.
In case the outward transport of the angular momentum by the shear
is compensated by an inward transport due to the meridional circulation,
we obtain the local conservation of the 
angular momentum. We call this solution the \emph{stationary solution}.
In this case, $U(r)$ is given by (cf. Zahn \cite{Za92})

\begin{equation}
U(r)= - \frac{5 D}{\Omega} \frac{\partial \Omega}{\partial r} \; .
\end{equation}

\noindent
The full solution of Eq.~(3) taking into account $U(r)$ 
and $D$ gives the 
\emph{non--stationary solution} of the problem. In this case, 
$\Omega (r)$ evolves as a result of the various transport processes,
according to their appropriate timescales, and in turn 
differential rotation influences the various above processes.
This  produces a feedback and,  thus, a self--consistent
solution for the evolution of $\Omega (r)$ has to be found.

The transport of angular momentum by circulation 
has often been treated as a diffusion process
(Endal and Sofia \cite{ES76}; Pinsonneault et al. \cite{Pin89};
Heger et al. \cite{He00}). From Eq.~(3), we see
that the term with $U$ (advection)
is functio\-nally not the same as the term with $D$ (diffusion).
Physically advection and diffusion are quite different:
diffusion brings a quantity  from where there is a lot to other
places where there is little. This is not necessarily the case
for advection. A circulation with
a positive value of $U(r)$, i.e.\ rising 
along the polar axis and descending at the equator, 
is as a matter of fact making an inward transport
of angular momentum. Thus, we see that when this process is treated as a
diffusion, like a function of $\frac{\partial \Omega}{\partial r}$, even
the sign of the effect may be wrong.

The expression of $U(r)$ given below (Eq.~12) involves derivatives up to the
third order, thus Eq.~(3) is of the fourth order, which makes the
system very difficult to solve numerically. 
In practice, we have applied a Henyey scheme to make the
calculations. Eq. (3) also implies four boundary conditions.
At the stellar surface, we take (cf. 
Talon et al. \cite{Ta97}; Denissenkov et al. \cite{Den99})

\begin{eqnarray}
\frac{\partial \Omega}{\partial r} = 0  \;\;\;\;
\mathrm{and}  \;\;\; \; U(r) =0
\end{eqnarray}

\noindent and at the edge of the core we have

\begin{eqnarray}
\frac{\partial \Omega}{\partial r} = 0  \; \; \; \;
\mathrm{and}  \; \; \; \;  \Omega(r) = \Omega_{\mathrm{core}.}
\end{eqnarray}

\noindent We  assume that 
the mass lost by stellar winds is just embarking its own angular momentum.
This means that we ignore any possible magnetic coupling, as it 
occurs in low mass stars.
This is not unreasonable in view of the negative results 
about the detection of magnetic fields in massive stars (Mathys \cite{Math99}).
It is interesting to mention here, that 
in case of no viscous, nor
magnetic coupling at the stellar surface, {\it i.e.} with the boundary
conditions (6),
the integration of Eq. (3) gives for an external shell of mass $\Delta M$ (Maeder 1999, paper IV)

\begin{eqnarray}
\Delta M {d \over dt} (\Omega r^2)=-{4\pi \over 5} \rho r^4 \Omega U(r).
\end{eqnarray}

\noindent This equation is valid provided the stellar winds are spherically symmetric (see paper IV),
an assumption we do in this work.
When the surface velocity approches the critical velocity, it is likely that there are  anisotropies of the mass
loss rates (polar ejection or formation of an equatorial ring)
and thus the surface condition should be modified according to 
the prescriptions of  Maeder (\cite{Mae99}). For now, these effects are not 
included in these models. Their neglect should not affect too much the results
presented here since the critical velocity is reached only in
some rare circumstances.

\subsection{Mixing and transport of the chemical elements}
A diffusion--advection equation like Eq.~(3) should normally 
be used to express the 
transport of chemical elements. However, if the
horizontal  component of the turbulent diffusion $D_{\rm{h}}$
is large,
the vertical advection of the elements  can be treated as 
 a simple diffusion
(Chaboyer and Zahn \cite{Cha92}) with a diffusion coefficient
$D_{\rm eff}$. As emphasized by Chaboyer and Zahn, this does
not apply to the transport of the angular momentum. $D_{\rm eff}$ is
given by

\begin{equation}
D_{\rm eff} = \frac{\mid rU(r) \mid^2}{30 D_h} \; ,
\end{equation} 

\noindent where $D_{\rm{h}}$ is the coefficient of horizontal
turbulence, for which
the estimate is

\begin{equation}
 D_{\mathrm{h}} \simeq |r U (r)|
\end{equation}

\noindent
according to Zahn (1992). Eq.~(8) 
expresses that the vertical
advection of chemical elements is severely inhibited by the
strong horizontal turbulence characterized by $D_{\rm{h}}$. 
 Thus, the change of the mass fraction $X_i$ of the chemical species 
$i$ is simply

\begin{eqnarray}
\lefteqn{ \left( \frac{dX_i}{dt} \right)_{M_r} = } \nonumber \\
&& \left(\frac{\partial  }{\partial M_r} \right)_t
\left[ (4\pi r^2 \rho)^2 D_{\rm mix} \left( \frac{\partial X_i}
{\partial M_r}\right)_t
\right] +  \left(\frac{d X_i}{dt} \right)_{\rm nucl} .
\end{eqnarray}

\noindent The second term on the right accounts for
 composition changes due
to nuclear reactions. The coefficient $D_{\rm mix}$ is the sum 
$D_{\rm mix} = D_{\rm shear}+D_{\rm eff}$
 and 
$D_{\rm eff}$ is given by Eq.~(9). 
The characteristic time for the mixing
of chemical elements is therefore
$t_{\rm mix} \simeq \frac{R^2}{D_{\rm mix}} $
and is not given by  $t_{\rm circ} \simeq \frac{R}{U}$,
as has been generally considered (Schwarzschild 1958).
This makes the mixing of the chemical elements much slower,
since $D_{\rm eff}$ is very much reduced. 
In this context, we recall that several
authors have  reduced  by large factors, up to 30 or 100,
the coefficient for the transport
of the chemical elements, with respect to the
transport of the angular momentum, 
in order to better fit the observed 
surface compositions (cf. Heger et al. \cite{He00}). This
reduction of the diffusion of the chemical 
elements is no longer
necessary with the more appropriate expression of $D_{\rm eff}$ given here.

When the effects of the shear
and of the meridional circulation compensate each other for the
transport of the angular momentum (\emph{stationary solution}, see Sect.~2.5), the value of $U$
entering the expression for $D_{\rm eff}$ is given by Eq. (5).

\subsection{Meridional circulation}
Meridional circulation is an essential mixing mechanism in rotating
stars and there is a considerable litterature on the subject
(see ref. in Tassoul \cite{Tass90}). 
The velocity of the meridional circulation in the case of shellular 
rotation was derived by Zahn (\cite{Za92}).
The effects of the vertical
$\mu$--gradient  $\nabla_{\mu}$  and of the horizontal 
turbulence  on meridional circulation are very important
and   they were taken into account by Maeder and Zahn  
(\cite{MZ98}).  
Contrarily to the conclusions of  previous works (e.g. Mestel \cite{Mes65};
Kippenhahn and Weigert  \cite{KippW90}; Vauclair \cite{Vau99}), 
the $\mu$--gradients  were shown not to introduce a
velocity threshold for the occurence of the meridional circulation,
but to progressively reduce
the circulation when $\nabla_{\mu}$ increases.
 The expression
by Maeder and Zahn (\cite{MZ98}) is

\begin{eqnarray}
\lefteqn{U(r) =  \frac{P}{{\rho} {g} C_{\!P} {T}
\, [\nabla_{\rm ad} - \nabla +
 (\varphi/\delta) \nabla_{\mu}] } } \nonumber  \\[2mm] 
 && \ \ \ \ \ \ \ \ \ \left\{  \frac{L}{M_\star} (E_\Omega + E_\mu) \right\} \; .
\end{eqnarray} 

\noindent
$P$ is the pressure, $C_P$ the specific heat, 
$E_{\Omega}$ and $E_{\mu}$ are  terms depending on the $\Omega$--
and $\mu$--distributions respectively, up to the third order derivatives
and on various thermodynamic quantities (see details in Maeder and Zahn, 
\cite{MZ98}).
The term  $\nabla_{\mu}$ in Eq.~(12) results from the
vertical chemical gradient and from the coupling between the
horizontal and vertical $\mu$--gradients due to the 
horizontal turbulence. This term $\nabla_{\mu}$ may be
one or two orders of magnitude larger than $\nabla_{\rm ad}-\nabla$
in some layers,
so that $U (r)$ may be reduced by the same ratio. This is one of the
important differences introduced by the work by
Maeder and Zahn (\cite{MZ98}). Another difference is that
the classical solution usually predicts an infinite
velocity at the interface between a radiative and a semiconvective
zone with an inverse circulation in the semiconvective
zone. Expression (12)
gives a continuity of the solution with no change of 
sign from semiconvective to radiative regions. Finally, 
we recall that in a stationary situation, $U(r)$ is given by Eq.~(5), as seen above.

\subsection{Shear turbulence and mixing}
In a radiative zone, shear due to differential rotation is likely
to be a most efficient mixing process. Indeed shear instability 
grows on a dynamical
timescale that is of the order of the rotation period 
(Zahn \cite{Za92}).  
The usual criterion for shear instability
is the Richardson criterion, which compares the balance
between the restoring force of the density gradient
and the excess energy present in the differentially rotating layers,

\begin{equation}
Ri = \frac{N^{2}_{\mathrm{ad}}}{(0.8836\ \Omega\frac {d\ln\Omega}{d\ln r})^2} < \frac {1}{4},
\end{equation}

\noindent
where we have taken the average over an isobar,
$r$ is the radius and
$N_{\rm{ad}}$ the Brunt-V\"ais\"al\"a frequency given by

\begin{eqnarray}
N^2_{\rm{ad}} = \frac{g \delta}{H_{P}} \left[ \frac{\varphi}
{\delta} \nabla_{\mu} + \nabla_{ad} - \nabla_{\rm{rad}} \right].
\end{eqnarray}

\noindent
When thermal dissipation is 
significant, the restoring force of buoyancy is reduced and
the instability occurs more easily,
its timescale is however longer, being the thermal timescale.
This case is  referred to as ``secular shear
instability''.  The criterion for low Peclet numbers $Pe$
(i.e. of large thermal dissipation, see below) has been 
considered by Zahn (\cite{Za74}), while the cases of
general Peclet numbers  $Pe$ have been considered by
Maeder (\cite{Mae95b}), Maeder and Meynet (\cite{MM96}), 
who give 

\begin{eqnarray}
\lefteqn{
Ri = \frac{g \delta}{(0.8836\ \Omega \frac{d\ln \Omega}{d\ln r})^{2} H_{P}}
}
 \nonumber  \\[2mm] 
 &&  \ \ \ \ \ \ \ \  \left[
\frac{\Gamma}{\Gamma +1} (\nabla_{ad} -\nabla) +
\frac{\varphi}{\delta} \nabla_{\mu} \right] < \frac{1}{4}
\end{eqnarray}

\noindent
The quantity $\Gamma =  Pe/6 $, where the Peclet number $Pe$
is the ratio of the thermal cooling time to the 
dynamical time, i.e.  $Pe = \frac{v \ell}{K}$ 
where $v$ and $\ell$ are the 
characteristic velocity and  length scales, and $K = (4acT^3)/
(3 C_P \kappa \rho^2 )$ is the thermal diffusivity. A discussion
of shear--driven turbulence by
Canuto (\cite{Ca98}) suggests that the limiting   $Ri$ number may
be larger than $\frac{1}{4}$. 

To account for shear transport and diffusion in Eqs.~(3) and (11),
we need a diffusion coefficient. Amazingly, a great variety of
coefficients  $D_{\mathrm{shear}} = \frac{1}{3} v \ell$
 have been derived and applied, for example:

 -- 1. Endal and Sofia (\cite{ES78})
apply the Reynolds and the Richardson criterion 
by Zahn (\cite{Za74}). They estimate $D_{\mathrm{shear}}$
from the product of the velocity scale height of the shear
flow and of the turbulent velocities of cells at the edge of
Reynolds critical number.

-- 2. Pinsonneault et al. (\cite{Pin89})
notice that the amount of differential rotation permitted by the
secular shears is proportional to a critical number, which they treat as
an adjustable parameter. To account for the effects of the $\mu$--gradient
in mixing, of the loss of angular momentum, etc\dots\ 
they introduce several adjustable
parameters in the equations for the
transport of the chemical elements and of the angular momentum.

-- 3. Chaboyer et al. (\cite{Cha95a})
use a coefficient derived from the velocity and path length from
Zahn (\cite{Za74}). Following Pinsonneault et al. (\cite{Pin89}),
they also introduce two adjustable parameters for adjusting the transports
of chemical elements and angular momentum respectively. We notice
that thanks to the reduction of the diffusion produced by
the horizontal turbulence (cf. Eq.~9 above), it is no longer
necessary to arbitrarily reduce the vertical transport
of the chemical elements.

-- 4. Zahn (\cite{Za92}) defines the diffusion coefficient cor\-responding
to the eddies which have the largest $Pe$ number so that the
Richardson criterion is just marginally satisfied. However, the
effects of the vertical $\mu$--gradient are not accounted for and
 the expression only applies to low Peclet numbers.

-- 5. The same has been done by Maeder and Meynet (\cite{MM96}),
who considers also the effect of the vertical $\mu$--gradient, the case of general
Peclet numbers and, 
in addition they account for the coupling due to the
fact that the shear also modifies the local thermal gradient.
This coefficient has been used by Meynet and Maeder (\cite{MM97})
and by Denissenkov et al. (\cite{Den99}).

-- 6. The comparisons  of model results and observations of
surface abundances have led many authors to conclude that the 
$\mu$--gradients  appear to inhibit  the shear mixing
too much with respect to what is required by the observations
(Chaboyer et al. \cite{Cha95a}b; 
Meynet and Maeder \cite{MM97}; Heger et al. \cite{He00}).
Namely, the observations of O--type stars (Herrero et al. 
\cite{He92}, \cite{He99}) show much more He-- and N--enrichments
than predicted by the models which apply Richardon's
criterion with the $\nabla_{\mu}$ term.
Thus, instead of
using a gradient $\nabla_{\mu}$ in the criterion for shear mixing,
 Chaboyer et al. (\cite{Cha95a}) and Heger et al. (\cite{He00})
 write  $f_{\mu} \nabla_{\mu}$ with a factor $f_{\mu} = 0.05$ 
or even  smaller.
This procedure is not satisfactory since it only accounts for a small fraction
of the existing $\mu$--gradients in stars.
The problem is that the models 
depend at least as much (if not more) on $f_{\mu}$ than on rotation, {\it i.e.} a change
of $f_{\mu}$ in the allowed range (between 0 and 1) produces as important effects
as a change of the initial rotational velocity.
This situation has led to two other more physical approaches discussed
below. Also Heger et al. (2000) introduce another factor $f_c$ to adjust
the ratio of the transport of the angular momentum and of the
chemical elements like Pinsonneault et al. (1989).

-- 7. Talon and Zahn (\cite{TZ97}) account for the horizontal
turbulence, which has a coefficient $D_{\rm{h}}$ 
and which weakens the restoring
force of the gradient of $\mu$ in the usual Richardson criterion.
This  allows some mixing of the chemical elements to occur 
as required by the observations (Talon et al. \cite{Ta97}).

-- 8.
Indeed,  around the convective core in the region
 where the $\mu$--gradient inhibits mixing,
there is anyway some turbulence due to both the  horizontal
turbulence and to the  semiconvective 
instability, which is  generally present in massive stars. 
This situation
has led  to the hypothesis  (Maeder \cite{Mae97})
that the  excess energy in the shear,
or a fraction $\alpha$ of it of the order of unity, 
is degraded by turbulence on the local thermal timescale.
This progressively changes the entropy gradient 
and consequently the $\mu$--gradient. This hypothesis
leads to  a diffusion coefficient $D_{\rm shear}$ given by

\begin{eqnarray}
D_{\rm shear} = 4 \frac{K}{N^{2}_{\rm{ad}}} \left[ 
\frac{1}{4} \alpha \left(0.8836\  \Omega \frac{d\ln\Omega}{d\ln r}
\right)^2 - (\nabla^{\prime} -\nabla) \right].
\end{eqnarray} 

\noindent
The term $\nabla^{\prime} -\nabla$ in Eq. (16) expresses either
the stabilizing effect of the thermal gradients in radiative zones or
its destabilizing effect in semiconvective zones (if any). 
When the shear is negligible, 
$D_{\rm shear}$ tends towards the diffusion coefficient for semiconvection
by Langer et al. (\cite{La83}) in semiconvective zones.
When the thermal losses are large ($\nabla^{\prime} =\nabla$), it tends towards the value

\begin{equation}
D_{\rm shear} = \alpha (K/N^2_{\rm{ad}}) \left(0.8836\  \Omega \frac{d\ln\Omega}{d\ln r}\right)^2 ,
\end{equation}

\noindent
given by Zahn (\cite{Za92}).
Eq.~(16) is completed by 
the three following equations expressing the thermal effects 
(Maeder \cite{Mae97})

\begin{eqnarray}
D_{\rm shear} = 2 K \Gamma \;\;\;\;\;\;\;\; \nabla=\frac{\nabla_{rad}+
(\frac{6 \Gamma^2}{1+\Gamma}) \nabla_{ad}}{1+(\frac{6 \Gamma^2}{1+\Gamma})},
\end{eqnarray}

\begin{eqnarray}
\nabla^{\prime} -\nabla = \frac{\Gamma}{\Gamma +1} (\nabla_
{\mathrm{ad}} - \nabla).
\end{eqnarray}

\noindent
The system of 4 equations given by Eqs. (16), (18) and (19) 
form a coupled
 system  with 4 unknown quantities
$D$, $\Gamma$, $\nabla$ and $\nabla^{\prime}$. 
The system is of the third degree in $\Gamma$.  
When it is solved   numerically,
we find  that as a matter of fact
the thermal losses in the shears are rather large in massive
stars  and thus that the Peclet number $Pe$ is very small
(of the order of 10$^{-3}$  to 10$^{-4}$, see Sect. 4.2).
For very low Peclet number
 $Pe =6 \Gamma$, the differences $(\nabla^{\prime} -\nabla)$ are 
also very small as shown by Eq.~(19).
Thus, we conclude that Eq.~(16) is essentially equivalent,
at least in massive stars, to the original Eq. (17) above, as
given by Zahn (\cite{Za92}). We may suspect that this is not
 necessarily true in low and intermediate mass stars since there the $Pe$ number may
be larger. Of course, 
the Reynolds condition 
$D_{\rm shear} \geq \frac{1}{3} \nu Re_c$ 
must be satisfied in order that  the medium is turbulent. The quantity
$\nu$ is the  total viscosity (radiative + mole\-cular) and 
 $Re_c$  the critical Reynolds number estimated to be
around 10 (cf. Denissenkov et al. \cite{Den99}; Zahn \cite{Za92}).
 The numerical results in Sect. 4  will 
show the values of the various parameters
and also indicate that the conditions
 for the occurence of turbulence are satisfied.

The physical treatment around the core also depends on the choice
of the criterion for convection. In current literature, there are at least three basic sets of assumptions: a) Ledoux criterion, which leads to small cores, b)
Schwarzschild's criterion, which gives ``medium size'' cores, c) 
Schwarzschild's criterion and overshooting, which gives large cores. Here, we
choose to apply the intermediate solution b), which means that the above equation 16 is only applied in fully radiative regions. Some consequences
of this choice of convective criterion are discussed in Sect. 3 below.

\subsection{Initial compositions, opacities, nuclear reactions and other model ingredients}

For purpose of comparison, we adopted here the same physical ingredients as for
the solar metallicity models computed by Meynet et al. (\cite{me94}). The only
exceptions, apart from the inclusion of the effects induced by rotation described
above, are the following: we use the Schwarzschild criterion for convection without
overshooting and the mass loss rates are as indicated in Sect. 2.4.

\section{The non--rotating stellar models}

For comparison purpose, we have computed non--rotating stellar models with
exactly the same physical ingredients as the rotating ones. The 
corresponding evolutionary tracks and lifetimes are presented in Fig.~8
and Table~1. 
These stellar models 
have similar properties as older ones computed with
the Schwarzschild criterion for convection. For instance, our non--rotating 9
M$_\odot$ stellar model  
has similar H-- and He--burning lifetimes as the 
9 M$_\odot$ model computed with the Schwarzschild criterion by Bertelli et al. (1985).
Our models also well agree with more
recent computations. Indeed our MS lifetimes are similar within about
two percents to the ones obtained by Heger et al. (2000)
for their non--rotating stellar models.
These authors used the Ledoux criterion with semiconvection. However,
since during the MS phase the convective core mass decreases, one expects,
for this phase, only small differences between the models computed with the Ledoux and the Schwarzschild criterion. These comparisons show that the present non--rotating stellar
models well agree with results obtained with different stellar evolutionary codes. 

Many papers (e.g. Bertelli et al. 1985; Maeder \& Meynet 1989; 
Chin \& Stothers 1990;
Langer \& Maeder 1995; Canuto 2000) have discussed
the effects of different criteria for convection on massive star evolution.
Since our main aim in this paper is to emphasize the effects of rotation, we shall restrict ourselves to briefly mention some differences with previous grids of models computed by the Geneva group.

In the present non--rotating stellar models, computed without overshooting, the convective
core masses are smaller than in the models by Schaller et al. (1992) and Meynet et al. (1994). 
As a  consequence, the stars evolve at smaller luminosities.
The turn--off point at the end of the MS occurs for
higher T$_{\rm eff}$ and lower luminosities, reducing the extent of
the MS width.
The MS lifetimes are reduced. For initial masses below about 25 M$_\odot$,
the He--burning lifetime is increased implying an increase of
the ratio $t_{\rm He}/t_{\rm H}$ of the lifetimes in the He-- and
H--burning phases. For the higher initial masses, the effects of the stellar
winds become very important and dominate the 
effects due to a change of the criterion for convection.

As a numerical example, in our non--rotating 20 M$_\odot$ stellar model,
the convective core mass is reduced by 1--1.2 M$_\odot$ during the whole H--burning phase compared to its value given by Meynet et al. (1994). 
At the end of the MS phase, the mass of the helium core has decreased
by $\sim$ 20\% with respect to its value in models
with a moderate overshoot. The position of the turn--off point has a $\log L$/L$_\odot$ decreased by
0.05 dex and a $\log T_{\rm eff}$ increased by 0.04 dex. The MS lifetime is
decreased by about 10\%, the ratio $t_{\rm He}/t_{\rm H}$ passes from 10\%
in the models of Meynet et al. (1994) to 14\% in the present case.

\section{The evolution of the internal rotation law $\Omega(r)$}

\subsection{The initial convergence of the rotation law}

Let us first mention that even in some recent works the 
assumption of solid body rotation is often considered. However, 
it is much more physical to examine the evolution of 
$\Omega(r)$ resulting from the transport of angular momentum
by shears, meridional circulation and convection, from the effects of central
contraction and envelope expansion and from the losses of
 angular momentum  by stellar winds. In particular,
the large losses of angular momentum at the surface lead to a redistribution of the
angular momentum in the interior by meridional circulation, shear
turbulent diffusion and convection.
The whole problem is self--consistent, since the meridional circulation 
and shear transport depend in turn on the degree of differential
rotation.Thus, the full system of equations
has to be solved in order to provide the internal evolution of $\Omega$.

\begin{figure}[tb]
  \resizebox{\hsize}{!}{\includegraphics{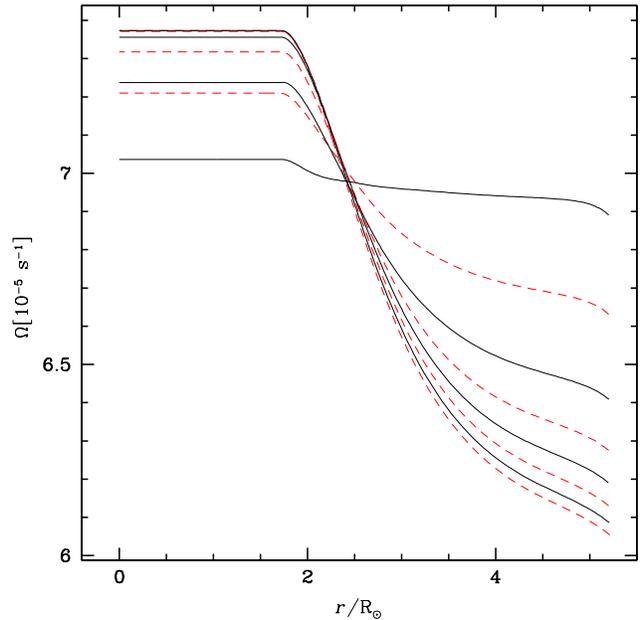}}
  \caption{Initial evolution of the angular velocity $\Omega$ 
 as a function of the distance to the center
in a 20 M$_\odot$ star with $v_{\rm ini}$ = 300 km s$^{-1}$.
Starting from a flat profile, the solutions rapidly converge
towards an equilibrium solution.
The time interval between two consecutive curves is 19\,200 years.
}
  \label{fconvo}
\end{figure}

\begin{figure}[tb]
  \resizebox{\hsize}{!}{\includegraphics{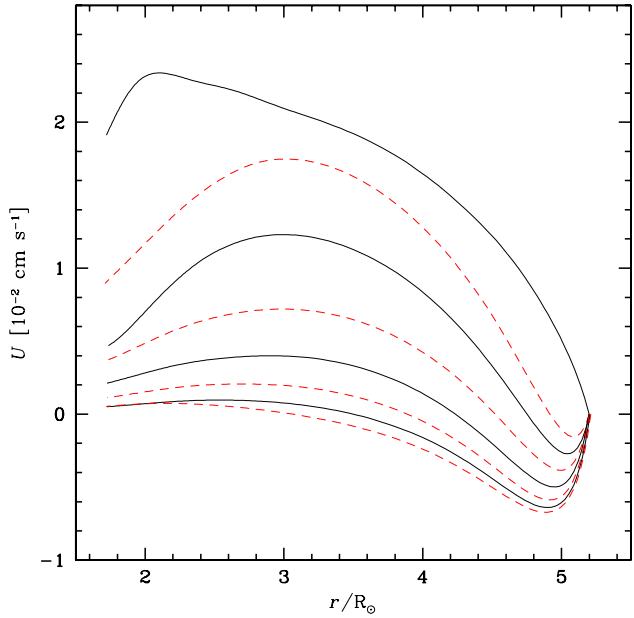}}
  \caption{Same as Fig.~\ref{fconvo} for $U(r)$ the vertical 
component of the meri\-dional circulation velocity (cf. Eq.~4). 
Time proceeds from top to bottom.
}
  \label{fconvu}
\end{figure}

Fig.~\ref{fconvo} shows the results for the initial convergence on the
zero age main sequence of 
$\Omega(r)$ in a 20 M$_{\odot}$ model with solar composition.
 The initial model on the zero--age main sequence 
is supposed to rotate uniformly and with an
initial surface velocity $v_{\rm ini}$ of 300 km s$^{-1}$. Very short time steps
of the order of 1920 yr are 
taken in the initial calculations for obtaining the internal
equilibrium rotation. The results are shown in Fig.~\ref{fconvo}
for every 10$^{\rm{th}}$ model, i.e. with time intervals of 19\,200 yr.
 We see the very fast initial changes  of $\Omega(r)$,
mainly due to the meridional circulation. The  circulation
velocity, which is very large, is also  positive
everywhere at the beginning (cf. Fig.~\ref{fconvu}).
This  means that the circulation rises along the polar axis
and goes down at the equator, thus bringing angular momentum inwards.
As a consequence the angular velocity of the core increases, 
while that in the envelope goes down. It is important
 to realize that here these fast changes of the rotation law 
are not due, 
as later in the evolution, to core contraction and envelope 
expansion. 

We see that the changes are big at the beginning and then smaller.
The profiles of $\Omega(r)$ rapidly converges towards
an equilibrium--profile in a time of about 1 to
2\% of the MS lifetime $t_{\rm MS}$. This is in agreement with
the results for 10 and 30 M$_{\odot}$ stars by Denissenkov et al.
(\cite{Den99}).
As noted by these authors, the timescale for the 
adjustment of the rotation law, i.e.  $t_{\rm circ} \simeq R/U$, 
 is short with respect to 
the MS lifetime for any significant rotation.  Later on during the bulk 
of MS evolution,
the profiles $\Omega(r)$ will evolve more slowly.
As emphasized in Sect. 2.6,
this timescale $t_{\rm circ}$ is not characteristic 
for the mixing of chemical elements.
We also notice that the degree of differential
rotation in the equilibrium profile is rather modest (cf. Zahn, 
\cite{Za92}).

Fig.~\ref{fconvu} shows the corresponding initial evolution of the 
vertical component $U(r)$ of the velocity of the meridional circulation
in the  20 M$_{\odot}$ model. As seen above, 
initially  this velocity is large and positive everywhere, which
explains the fast evolution of $\Omega(r)$ in Fig.~\ref{fconvo}. Then, $U(r)$
decreases and becomes negative in the very external layers.
The physical reason for that is the so called  Gratton--\"Opik term
of the form $- \frac{\Omega^2}{2 \pi G \rho}$ which is contained in
the expression of $E_{\Omega}$ in Eq.~(12). When the density is very low,
as in the outer regions, this negative term becomes important and 
produces an inverse circulation.
 This
means that the circulation has two big cells, an internal cell rising along
the polar axis and an external one descending at the pole.
The velocity $U(r)$ also converges towards an equilibrium 
distribution, characterized by small velocities $U(r)$, a result 
in agreement with Denissenkov et al. (\cite{Den99}).

\begin{figure}[tb]
  \resizebox{\hsize}{!}{\includegraphics{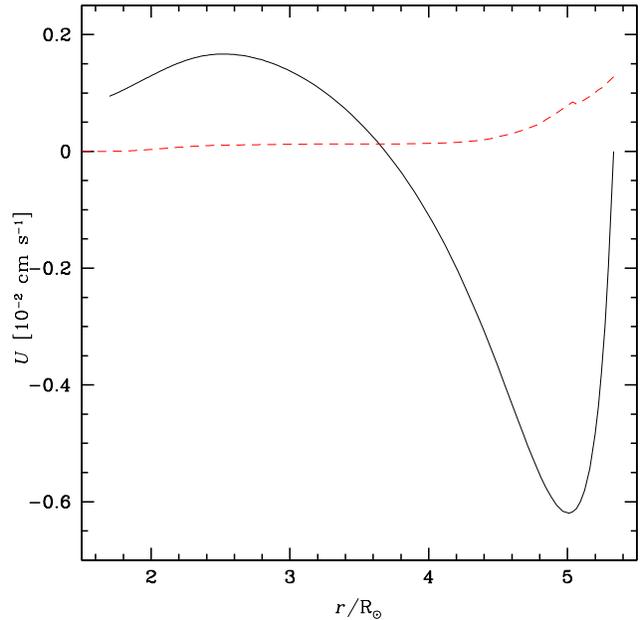}}
  \caption{Non--stationary (continuous line) and stationary (broken line) solutions
for $U(r)$ the vertical component 
of the velocity of the meridional circulation (cf. Eq.~4) as a function of the distance to the center
in the  20 M$_{\odot}$ model at the end of the initial convergence.}
  \label{fdifu}
\end{figure}

It is interesting to compare the stationary solution of
$U(r)$ given by Eq.~(5) and the non--stationary solution
given by Eq.~(12). This is done in Fig.~\ref{fdifu}, where the curve
in continuous line
corresponds to the asymptotic distribution
reached in the non--stationary regime illustrated in Fig.~\ref{fconvu}. 
This curve shows
positive values of $U(r)$ in the inner regions and negative values in the
outer ones. On the contrary, the solution (broken line)
obtained in the stationary approximation given by Eq.~(5) is
always positive. This is a logical consequence of 
the  approximation made: 
as $\Omega(r)$ decreases outwards, only positive values of $U(r)$
are possible. Said in other words, the outward transport by shears
can only be compensated by the inward transport by meridional circulation, {\it i.e.}
by positive values of $U(r)$. 

However, in non--stationary models,
this compensation is not achieved since
the two curves in Fig.~\ref{fdifu} are very different. One
concludes that the stationary  approximation 
which has been used in some works (cf. Urpin et al. \cite{Urp96}) 
is  not satisfactory. 

The stationary solution can also be viewed as giving a velocity
which is the inverse of a velocity which could be associated to the diffusive transport
by shears.
From the comparison of the two curves in Fig.~\ref{fdifu}, one thus immediately
concludes that the meridional circulation is much more efficient
for the transport of the angular momentum than the shear instabilities.
The justification for this claim is the following one. In the stationary
situation, shear transport and meridional circulation compensate each
other and are thus of the same magnitude. The fact that $U(r)$ in the full non--stationary
case treatment is much larger than in the stationary case as shown in Fig.~3 implies that the effects of meridional
circulation on the transport of angular momentum are much larger 
than the effects of the shear.
This conclusion can also be deduced from the comparison of the timescales for the two
processes. In general one has that $t_{\rm circ}=\frac{R}{U} < t_{\rm shear}=
\frac {R^2} {D_{\rm shear}}$. For the chemical elements,
the transport by the shear instabilities is more efficient than
the transport by the meridional circulation as discussed in 
Sect. 2.6 where we have seen that $D_{\rm eff}$ is reduced by horizontal turbulence.

\subsection{Internal evolution of $\Omega(r)$}

\begin{figure}[tb]
  \resizebox{\hsize}{!}{\includegraphics{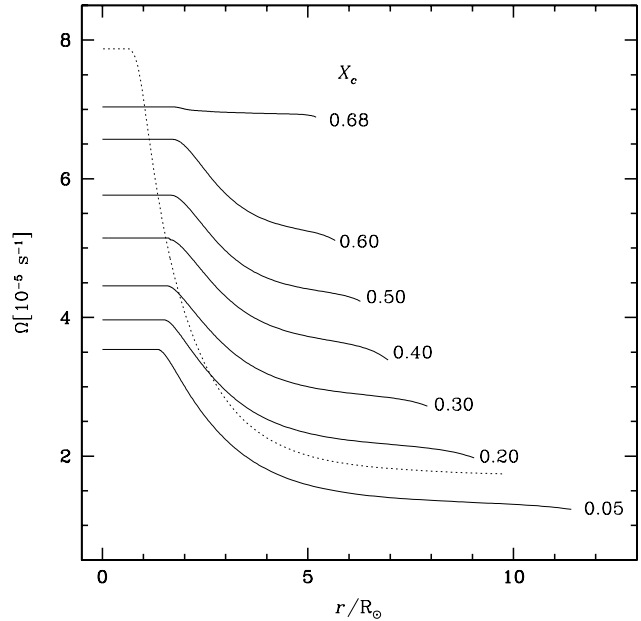}}
  \caption{Evolution of the angular velocity $\Omega$ 
 as a function of the distance to the center
in a 20 M$_\odot$ star with $v_{\rm ini}$ = 300 km s$^{-1}$. 
$X_c$ is the hydrogen mass fraction at the center.
The dotted line shows the profile when the He--core contracts at the end
of the H--burning phase.}
  \label{fom}
\end{figure}

The evolution of $\Omega(r)$  during the
MS evolution of a 20 M$_{\odot}$ star is shown in Fig.~\ref{fom}.
We notice that initially (i.e. for high values of the central H--content
$X_{\rm{c}}$), the degree of differential rotation is small,
then, it grows during the evolution. The ratio between the central and surface
value of $\Omega$ remains however small during the MS evolution. The
rotation rate does not vary by more than about a factor two throughout the star.
{\it Although this degree of differential rotation is not large, it plays an
essential role in the shear effects and the related transports of angular momentum and of chemical elements.} The same remark also applies to the effects
of differential rotation on the transport by meridional circulation, since
$U(r)$ critically depends on the derivatives of $\Omega$ in the star
(cf. Zahn 1992; Maeder \& Zahn 1998).

We notice a general  decrease of 
$\Omega(r)$, even in the convective core
which is contracting. The main reason is  mass loss 
at the stellar surface, which  removes a substantial
fraction of the total angular momentum. This makes 
 $\Omega(r)$  to decrease with time everywhere, because of the
internal transport mechanisms, which ensure some coupling of rotation.
The same behaviour was obtained by Langer (1998) in the case
of rigid rotation ({\it i.e.} in the case of maximum coupling).
In the present models,
shear transports the angular momentum outwards,
which reduces the core rotation, while circulation makes an inward 
transport in the deep interior and an outwards transport 
in the external region. This is responsible for the progressive flattening
of the curves above $r = 4 R_{\odot}$ in Fig.~\ref{fom}.

From the end of the MS evolution onwards, i.e. when $X_c \leq 0.05$, 
central contraction becomes faster and 
starts dominating the evolution of the angular momentum in the center.
The central
$\Omega$ grows quickly (cf. dotted curve in Fig.~\ref{fom})
and this will in principle 
continue in the further evolutionary phases
until core collapse. The evolution of the angular momentum 
after the H--burning phase ($X_c = 0$) is mainly dominated by the local 
conservation of the angular momentum, since the secular transport mechanisms
have little time to operate. This is the assumption we are making in the
present models. During these phases, the chemical elements continue to be rotationally mixed in the radiative layers
mainly through the effect of the shear.
In the further evolutionary phases,
the strong central contraction
will lead to very large central rotation, unless some 
other processes as
fast dynamical instabilities or magnetic braking  occur in these stages.

\begin{figure}[tb]
  \resizebox{\hsize}{!}{\includegraphics{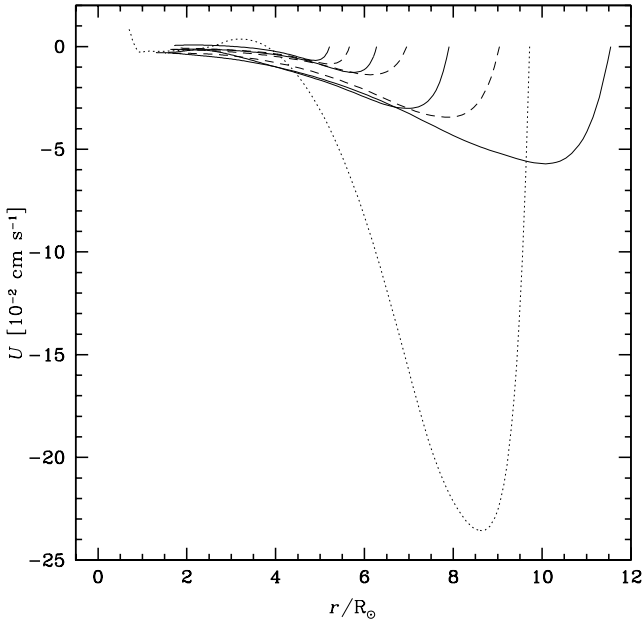}}
  \caption{Same as Fig.~\ref{fom} for $U(r)$.
}
  \label{fur}
\end{figure}

The evolution of $U(r)$ during MS evolution is shown in Fig.~\ref{fur}.
 The outer zone 
with inverse  circulation progressively deepens in radius 
during MS evolution due to the  Gratton--\"Opik term,
because a growing part of the outer layers has  lower densities.
Also, we notice that the velo\-cities $U(r)$ are small in general
(cf. Urpin et al. \cite{Urp96}) and tests have shown us that,
contrary to the classical result of the 
Eddington--Sweet circulation (cf. Mestel 1965), $U(r)$ depends rather little 
on the initial rotation.

The deepening
of the inverse circulation  has the consequence that the
stationary and non--stationary solutions differ more and
more as the evolution proceeds, since as said above no inverse
circulation is predicted by the stationary solution.
Thus we conclude that  the stationary
solutions  are  much too simplified. Also,
in Fig.~\ref{fur}, we see that the values of $U(r)$ become more negative
in the outer layers for the model at the end of the MS phase, when
central contraction occurs. Again, this effect is due to the Gratton--\"Opik term. The results of Fig.~5 also shows that the application of a
diffusive treatment to meridional circulation transport is unappropriate.
Paradoxically, the problem is less serious in regions which
exhibit an inverse circulation, since there the signs of the effects are at least the same in the two treatments (however even there the equations
for advection and diffusion are different).

\begin{figure}[tb]
  \resizebox{\hsize}{!}{\includegraphics{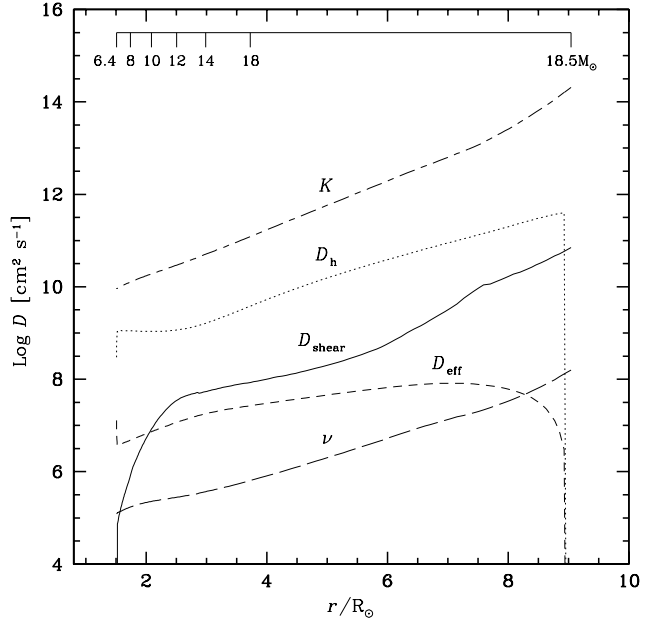}}
  \caption{Internal values  of $K$ the thermal diffusivity,
 $D_{\rm h}$ the coefficient of
horizontal turbulence, $D_{\rm shear}$ the shear diffusion coefficient,
$D_{\rm eff}$ the effective diffusivity (cf. Eq.~9) and $\nu$ the total
viscosity (radiative + molecular)
in the radiative envelope of a 20 M$_\odot$ star 
with an initial $v_{\rm ini}$ = 300 km s$^{-1}$. 
The lagrangian mass coordinate is given on the upper scale.
Here, the hydrogen mass fraction at the center $X_c = 0.20$.}
  \label{fdiff}
\end{figure}

The various diffusion coefficients inside a 20 
M$_{\odot}$ star  are shown in Fig. 6, when the hydrogen 
mass fraction $X_c$ at the center is  equal to 0.20.
We notice that in general $ K > D_h  > D_{\rm shear} >
D_{\rm eff} > \nu$. Following the thermal diffusivity $K$, 
the coefficient of horizontal 
diffusion $D_h$ is the second largest one,
 this is consistent and necessary for
 the validity of the assumption of shellular rotation (Zahn, \cite{Za92}).  
Since we have $D_{\rm{shear}} = 2 K \Gamma$ according to Eq.~(18),
Fig.~\ref{fdiff} shows that 
$\Gamma$ is typically of the order of 10$^{-3}$ to 10$^{-4}$. 
The coefficient 
$D_{\rm{eff}}$, expressing the transport of the chemical elements by the meridional
circulation with account of the reduction by the horizontal turbulence,
is only slightly smaller (by a factor of about 3)
than $D_{\rm{shear}}$ in the interior. One must be careful that
 $D_{\rm{eff}}$ concerns only the tranport of the elements and not that
of angular momentum, for which meridional circulation is the largest effect. 
Fig. 6 also shows the values of the total viscosity $\nu$,  and we
 notice that consistently the Prandtl number $\nu/K$  is of 
the order of 10$^{-5}$ to 10$^{-6}$.

\section{HR diagram, lifetimes and age estimates}

\subsection{The Main--Sequence evolution}



\begin{figure}[tb]
  \resizebox{\hsize}{!}{\includegraphics{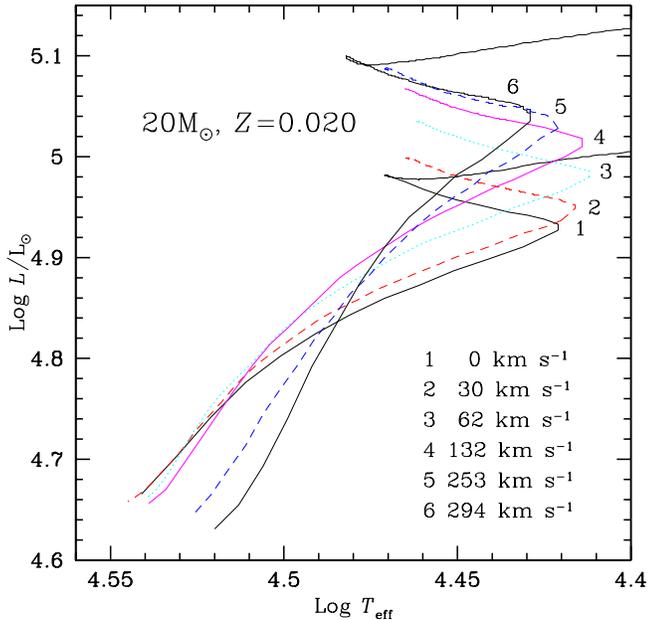}}
  \caption{Evolutionary tracks for rotating 20 M$_\odot$ models with different
initial velocities. 
The mean equatorial velocities $\overline{v}$ during the MS are indicated. 
The corresponding initial velocities
can be seen in Table~2.}
  \label{HRvingt}
\end{figure}

Evolutionary tracks of 20 M$_\odot$ models at solar metallicity for different
initial velocities are plotted on Fig.~\ref{HRvingt}. The $T_{\rm eff}$ for
a rotating star has been defined in Sect. 2.3.
On and near the ZAMS, rotation produces a small shift of the tracks
towards lower luminosities and $T_{\rm eff}$. This effect is due to both
atmospheric distorsions and to the lowering of the effective gravity 
(see e.g. Kippenhahn and Thomas \cite{KippTh70}; Maeder and Peytremann \cite{MP70}; 
Collins and Sonneborn \cite{co77}). At this stage the star
is still nearly homogeneous.
When evolution proceeds, the tracks with rotation become more luminous than for
no rotation.
This results from essentially two effects. On one side, rotational mixing
brings fresh H--fuel into the convective core, slowing down its decrease in mass
during the MS. For a given value of the central H--content,
the mass of the convective core in the rotating model
is therefore larger than in the non--rotating one and thus the stellar luminosity is higher
(Maeder \cite{ma87}; Talon et al. \cite{Ta97}; Heger et al. \cite{He00}).
As a numerical example,
in
the $v_{\rm ini}=$ 300 km s$^{-1}$ models the He--cores at the end of the MS are about 20\% more massive than in their
non--rotating counterparts. This is equal to the increase obtained by a
moderate overshooting (see Sect. 3).
On the other side, rotational mixing transports
helium and other H--burning products (essentially nitrogen)
into the radiative envelope. The He--enrichment lowers the opacity. This
contributes to the enhancement of the stellar luminosity and favours a blueward track.
Indeed, in
Fig.~\ref{HRvingt}, one sees that when the mean velocity on the MS becomes
larger than about 130 km s$^{-1}$, the $T_{\rm eff}$ at the end of the MS increases when rotation increases.

For sufficiently low velocities, rotation acts as a small
overshoot, extending the MS tracks towards lower $T_{\rm eff}$. This results from the fact
that, at sufficiently low rotation,
the effect of rotation on the convective core mass overcomes the effect of helium
diffusion in the envelope. Indeed, for small rotation, the time required for 
helium mixing in the whole radiative envelope is very long, 
while hydrogen just needs to diffuse 
through a small amount of mass to reach the convective core.

\begin{figure*}[tb]
  \resizebox{\hsize}{!}{\includegraphics[angle=-90]{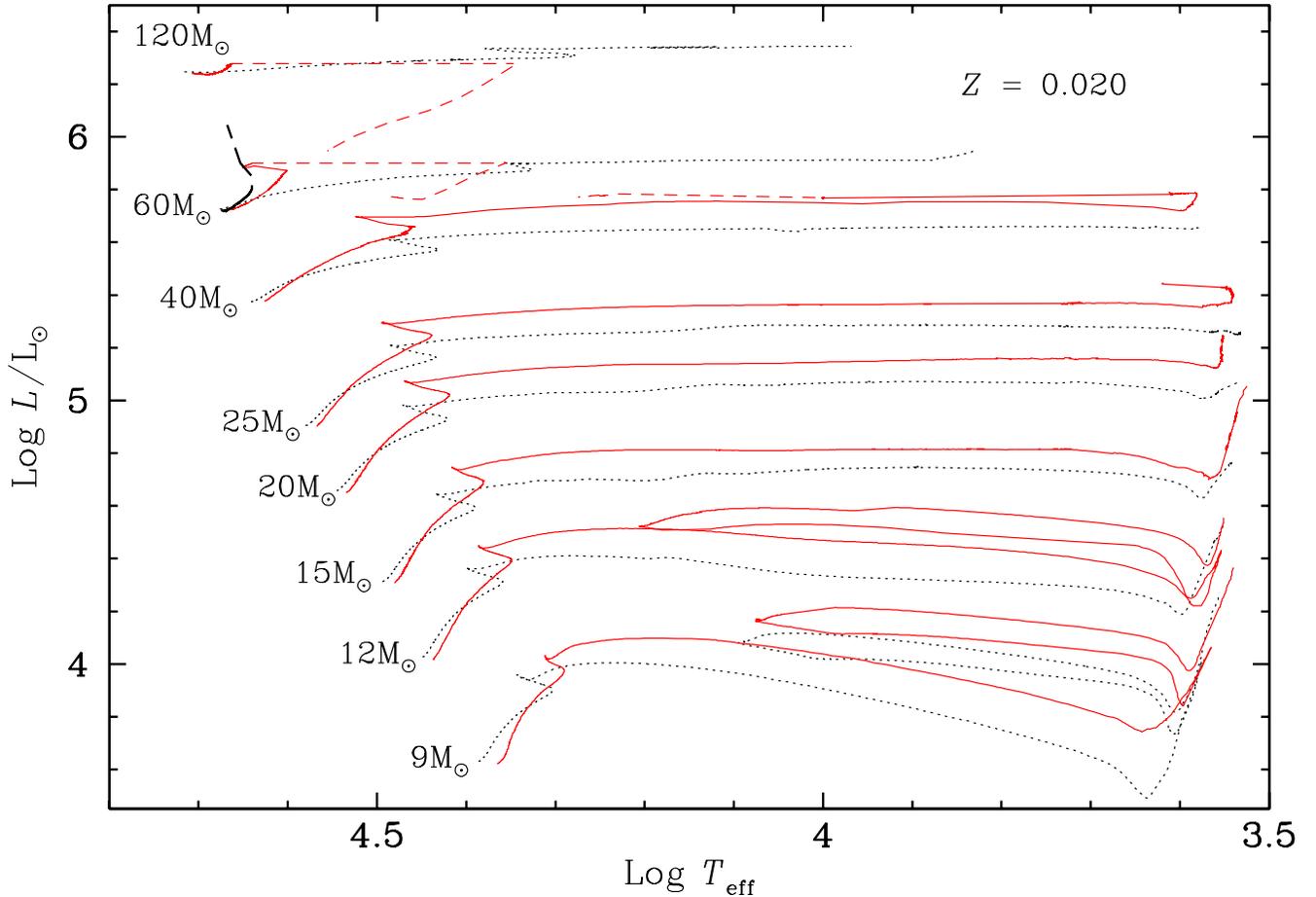}}
  \caption{Evolutionary tracks for non--rotating 
(dotted lines) and
rotating (continuous lines) models with solar metallicity. 
The rotating models
have an initial velocity $v_{\rm ini}$ of 300 km s$^{-1}$.
For purpose of clarity, only the first part of the tracks for the most massive stars (M $\ge 40$ M$_\odot$) is shown.
Portions of the evolution during 
the W--R phase for the rotating massive stars are indicated by
short--dashed lines. 
The long--dashed track for the 60 M$_\odot$ model corresponds to a very fast rotating star
($v_{\rm ini}$ $\sim 400$ km s$^{-1}$), which
follows a nearly homogeneous evolution. Only the beginning of its evolution is 
shown.}
  \label{HRge}
\end{figure*}

Since not all the stars have the same initial
rotational velocity, one expects a dispersion of the luminosities
at the end of the MS. For the 20 M$_\odot$ models shown on Fig.~\ref{HRvingt}
one sees a difference of $\sim 0.3$ in $M_{\rm bol}$ between
the luminosities of
the low and fast rotators at the end of the MS. 
Rotation induces also a scatter of the effective 
temperatures at the end of the MS. In reality,
the
dispersion results from both different initial velocities and also, for a given initial velocity,
from different angles between the axis of rotation and the line of sight. 
Indeed due to the von Zeipel theorem (\cite{vZ24}) the star
appears bluer seen pole--on than equator--on. When integrated
over the visible part of the star, the effects due to orientation can reach
a few tenths of a magnitude in luminosity and a few hundredths in $\log T_{\rm eff}$
(cf. Maeder and Peytreman 1970).

Figure~\ref{HRge} shows the evolutionary tracks of non--rotating and rotating
stellar models for initial masses between 9 and 120 M$_\odot$. For the rotating stellar models, 
the initial velocity is $v_{\rm ini}$ = 300 km s$^{-1}$.
There is little difference between tracks with 
$v_{\rm ini}$ = 200 or 300 km s$^{-1}$ (see also Talon et al. \cite{Ta97}).
If the effects behaved  like $v_{\rm ini}^2$, there 
would be  larger differences.
The present saturation effect occurs because outward transport of angular momentum by shears are larger when rotation is larger, also
a larger rotation produces 
more mass loss, which makes a larger reduction of rotation during the evolution. Let us note however that for some surface abundance ratios as
N/C or N/O (see Table~1), the increase from $v_{\rm ini}$ = 200 to 300 km s$^{-1}$
produces significant changes. Thus, the similarity of the evolutionary 
tracks does
not necessarily imply the similarity of the surface abundances for these
elements. 

Rotation reduces the MS width in the high mass range (M $\simlr 40$ M$_\odot$). 
Let us recall that
when the mass increases, the ratio of the diffusion timescale for the
chemical elements to the MS lifetime decreases (Maeder \cite{ma98}). As a consequence, starting
with the same $v_{\rm ini}$ on the ZAMS,
massive stars will be more mixed than low mass stars at an identical stage
of their evolution. This reduces the MS width since greater
chemical homogeneity makes the star bluer. 
Moreover, due to both rotational mixing and mass loss, their surface will be
rapidly enriched in H--burning products. These stars will therefore enter the 
Wolf--Rayet phase while they are still burning their hydrogen in their core. This again reduces the MS width.
For initial masses between 9 and 25 M$_\odot$, the MS shape is not much changed by rotation at least
for $v_{\rm ini} \le 300$ km s$^{-1}$.

\subsection{The post--Main-Sequence evolution}

The post--MS evolution of the most massive stars (M $\ge 40$ M$_\odot$) which become W--R stars
will be discussed in a forthcoming paper. We shall just mention one point of general interest
here: for low or moderate rotation, the
convective core shrinks  as usual during MS evolution,
while for high masses ($M \simgr 40$ M$_{\odot}$) and  large initial rotations 
($\frac{\Omega}{\Omega_{\rm crit}}
\geq 0.5 $), the convective core grows in mass  during 
evolution. This latter situation occurs in the fast rota\-ting 60 M$_\odot$ model
shown on Fig.~\ref{HRge}.
These behaviours, i.e.\ reduction or growth of the core,
determine  whether the star will follow 
respectively the usual redwards MS tracks in the HR diagram, 
or whether it will bifurcate to the blue (cf. Maeder \cite{ma87}; Langer \cite{la92})
towards the classical tracks of homogeneous evolution (Schwarzschild \cite{sc58})
and likely produce W--R stars.

The stars with initial masses between 15 and 25 M$_\odot$
become red supergiants (RSG). Rotation does not change qualitatively this behaviour but
accelerates the redwards evolution, especially for the 15 and 20 M$_\odot$ models. As a numerical example, for an initial
$v_{\rm ini}$ = 300 km s$^{-1}$, the model stars burn all their helium as red supergiants at $T_{\rm eff}$
below 4000 K, while
the non--rotating models spend a significant part of the He--burning phase in the blue part
of the HR diagram: for the non--rotating 15 and 20 M$_\odot$ models, respectively 25 and 20\% of the total He--burning
lifetime is spent at $\log T_{\rm eff} \ge 4.0$. The behaviour of the rotating models results 
mainly from the enhancement 
of the mass loss rates. This effect prevents the formation of a big intermediate convective zone
and therefore favours a rapid evolution toward the RSG phase 
(Stothers and Chin \cite{st79}; Maeder \cite{ma81}). 
Let us note that the dispersion of the initial rotational
velocities produces a mixing of the above behaviours.

Very interestingly, for the 12 M$_\odot$ model a blue loop appears when
rotation is included. This results from the higher luminosity of the
rotating model. The higher luminosity implies that the outer envelope is more extended, and is thus characterized
by lower temperatures and higher opacities at a given mass coordinate.
As a consequence,  in the rotating model during
the first dredge--up, the outer convective zone
proceeds much more deeply in mass than in the non--rotating star. 
Typically in the non--rotating model the 
minimum mass coordinate reached by the outer
convective zone is 6.6 M$_\odot$ while in the rotating model it is 2.6 M$_\odot$.
This prevents
temporarily the extension in mass of the He--core and enables
the apparition of a blue loop. Indeed the lower the mass of the He--core is, the lower
its gravitational potential. According to 
Lauterborn et al. (\cite{la71}, see also the discussion in Maeder and Meynet \cite{ma89}), 
a blue loop appears when the gravitational potential
of the core $\Phi_c$ is inferior to a critical potential $\Phi_{\rm crit}$ 
depending only on the actual mass of the star which is about the same
for the rotating and non--rotating model. This explains 
the appearance of a blue loop in the 12 M$_\odot$ rotating model.
For the 9 M$_\odot$ model, the minimum mass coordinate reached by the outer
convective zone is not much affected by rotation and the models with and without rotation present
very similar blue loops.

\begin{table*}
\caption{Properties of the stellar models at the end of the H--burning phase, at the blue supergiant (BSG) or LBV stage (see text) 
and at the end of the He--burning phase. The masses are in solar mass, the velocities in km s$^{-1}$, the lifetimes in million years and the abundances in mass fraction.} \label{tbl-1}
\begin{center}\scriptsize
\begin{tabular}{ccc|cccccc|cccc|cccccc}
\hline
    &         &         &         &         &     &      &      &      &     &      &      &      &         &         &     &      &      &       \\
M & $v_{\rm ini}$ & $\overline{v}$ & \multicolumn{6}{|c|}{End of H--burning}&\multicolumn{4}{|c|}{BSG or LBV} &\multicolumn{6}{|c}{End of He--burning} \\
    &     &    &     &         &     &      &      &      &     &      &      &      &         &         &     &      &      &       \\    
    &  &   & $t_H$ & M & $v$ & Y$_s$ & N/C & N/O & $v$ & Y$_s$ & N/C & N/O & $t_{He}$ & M & $v$ & Y$_s$ & N/C &  N/O \\
    &   &  &         &         &     &      &      &      &     &      &      &      &         &         &     &      &      &       \\
\hline
    &  &   &         &         &     &      &      &      &     &      &      &      &         &         &     &      &      &       \\
120 &  0 & 0 &  2.557 & 76.070 &   0 & 0.55 & 57.8 & 1.48 &   0 & 0.66 & 55.6 & 24.2 & 0.326  & 58.024 &  0  & 0.85 & 47.4 & 45.1  \\
    &  &   &         &         &     &      &      &      &     &      &      &      &         &         &     &      &      &       \\
    & 300 & 163 &  2.890 & 57.901 &  65 & 0.89 & 49.3 & 45.4 &  65 & 0.91 & 48.5 & 46.8 & 0.357  & 16.201 & 27  & 0.22 & 0    & 0     \\
    &   &  &         &         &     &      &      &      &     &      &      &      &         &         &     &      &      &       \\
 60 &  0 & 0 &  3.366 & 47.517 &   0 & 0.30 & 0.26 & 0.12 &   0 & 0.46 & 10.3 & 2.26 & 0.394  & 14.960 &  0  & 0.23 & 0    & 0     \\
    & 200 & 107 &  3.922 & 40.989 &  29 & 0.59 & 18.6 & 5.55 &     &      &      &      &         &         &     &      &      &       \\
    & 300& 168 &  4.128 & 25.066 &  29 & 0.90 & 49.6 & 30.2 &  49 & 0.93 & 49.6 & 34.8 & 0.423  & 11.697 & 46  & 0.39 & 0    & 0     \\
    & &     &         &         &     &      &      &      &     &      &      &      &         &         &     &      &      &       \\
 40 &  0&  0 &  4.155 & 34.761 &   0 & 0.30 & 0.25 & 0.12 &   0 & 0.30 & 0.25 & 0.12 & 0.473  & 12.565 &  0  & 0.98 & 41.4 & 47.4  \\
    & 200 & 114 &  4.936 & 31.871 &  38 & 0.43 & 4.29 & 1.49 &     &      &      &      &         &         &     &      &      &       \\
    & 300 & 172 &  5.105 & 30.898 & 104 & 0.47 & 5.33 & 1.85 &  15 & 0.48 & 5.39 & 1.87 & 0.462  & 11.872 & 119 & 0.35 & 0    & 0     \\
    &  &    &         &         &     &      &      &      &     &      &      &      &         &         &     &      &      &       \\
 25 & 0 &  0 &  5.928 & 23.213 &   0 & 0.30 & 0.25 & 0.12 &   0 & 0.30 & 0.25 & 0.12 & 0.737  & 18.912 &  0  & 0.44 & 3.95 & 1.22  \\
    & 200 & 125 &  7.114 & 22.089 &  77 & 0.34 & 1.40 & 0.52 &     &      &      &      &         &         &     &      &      &       \\
    & 300 & 183 &  7.442 & 21.640 & 154 & 0.37 & 2.07 & 0.72 &  90 & 0.38 & 2.15 & 0.75 & 0.718  & 11.657 & 0.4 & 0.65 & 36.4 & 3.40  \\
    &  &    &         &         &     &      &      &      &     &      &      &      &         &         &     &      &      &       \\
 20 & 0 &  0 &  7.350 & 19.019 &   0 & 0.30 & 0.25 & 0.12 &   0 & 0.30 & 0.25 & 0.12 & 1.032  & 17.043 &  0  & 0.38 & 2.02 & 0.70  \\
    & 200 & 132 &  8.901 & 18.324 &  94 & 0.32 & 1.01 & 0.38 &     &      &      &      &         &         &     &      &      &       \\
    & 300 & 197 &  9.309 & 18.020 & 167 & 0.35 & 1.77 & 0.58 &  34 & 0.35 & 1.80 & 0.59 & 0.871  & 14.605 & 0.3 & 0.48 & 5.70 & 1.40  \\
    &  &    &         &         &     &      &      &      &     &      &      &      &         &         &     &      &      &       \\
 15 & 0 &  0 & 10.214 & 14.631 &   0 & 0.30 & 0.25 & 0.12 &   0 & 0.30 & 0.25 & 0.12 & 1.506  & 13.728 &  0  & 0.34 & 1.30 & 0.45  \\
    & 200 & 145 & 12.316 & 14.365 & 142 & 0.31 & 0.69 & 0.26 &     &      &      &      &         &         &     &      &      &       \\
    & 300 & 209 & 12.917 & 14.260 & 226 & 0.32 & 1.36 & 0.43 &  60 & 0.33 & 1.43 & 0.45 & 1.482 & 12.589 & 31  & 0.44 & 4.69 & 1.06  \\
    &  &   &         &         &     &      &      &      &     &      &      &      &         &         &     &      &      &       \\
 12 &  0 & 0 & 13.929 & 11.926 &   0 & 0.30 & 0.25 & 0.12 &   0 & 0.30 & 0.25 & 0.12 & 2.368  & 11.100 &  0  & 0.32 & 1.07 & 0.36  \\
    & 200 & 150 & 16.069 & 11.867 & 141 & 0.30 & 0.60 & 0.23 &     &      &      &      &         &         &     &      &      &       \\
    & 300 & 217 & 16.797 & 11.828 & 242 & 0.31 & 1.07 & 0.34 & 139 & 0.41 & 4.37 & 0.91 & 2.503  & 10.873 & 1.2 & 0.41 & 4.42 & 0.91  \\
    &  &    &         &         &     &      &      &      &     &      &      &      &         &         &     &      &      &       \\
  9 & 0 &  0 & 22.054 &  8.991 &   0 & 0.30 & 0.25 & 0.12 &   0 & 0.32 & 1.32 & 0.42 & 3.728  &  8.875 &  0  & 0.32 & 1.32 & 0.42  \\
    & 200 & 153 & 25.862 &  8.982 & 158 & 0.30 & 0.41 & 0.17 &     &      &      &      &         &         &     &      &      &       \\
    & 300 & 235 & 26.737 &  8.977 & 266 & 0.31 & 0.86 & 0.29 &  98 & 0.37 & 3.59 & 0.73 & 3.997  &  8.770 & 2.4 & 0.37 & 3.59 & 0.73  \\
    &  &    &         &         &     &      &      &      &     &      &      &      &         &         &     &      &      &       \\
\hline
\end{tabular}
\end{center}

\end{table*}

\subsection{Masses and mass--luminosity relations}

When rotation increases, the actual masses at the end of both the MS and the He--burning phases
become smaller (cf. Tables~\ref{tbl-1} and \ref{tbl-2}).
Typically the quantity of mass lost by stellar winds during the MS is enhanced by
60--100\% in rotating models with $v_{\rm ini}$ = 200 and 300 km s$^{-1}$  respectively. For stars which do not go through a Wolf--Rayet phase,
the increase is due mainly to the direct effect of rotation on the mass loss rates (in the present models through
the formula proposed by Friend and Abbott \cite{Fr86}) and to the higher luminosities
reached by the tracks computed with rotation. The fact that rotation increases the lifetimes
also contributes to produce smaller final masses. 
For the most massive stars (M $\ge 60$ M$_\odot$), the present rotating models 
enter the Wolf--Rayet phase already during the
H--burning phase (see also Maeder 1987; Fliegner \& Langer 1995;
Meynet 1999, 2000b). This reduces significantly the mass at the end of the H--burning phase. 

As indicated in Sect. 5.1, the initial distribution of the rotational velocities
implies a dispersion of the luminosities at the end of the MS.
This effect introduces a significant scatter in the mass--luminosity
relation (Langer 1992; Meynet \cite{me98}), in the sense that fast rotators are
overluminous with respect to their actual masses.
This is especially true in the high mass star range in which the luminosity versus mass relation
flattens. 
This may explain some of the discrepancies
between the evolutionary masses and the direct
mass estimates in some
binaries (Penny et al. \cite{pe99}).

Let us end this section by saying a few words about the mass discrepancy problem (see e.g.
Herrero et al. \cite{He00}). For some stars, the evolutionary masses ({\it i.e.} determined from
the theoretical evolutionary tracks) are greater that the spectroscopically determined masses.
Interestingly,
according to
Herrero et al. (\cite{He20}), only the low gravity objects
present (if any) a mass discrepancy. Even if most of the
problem has collapsed and was shown 
to be a result of the proximity of O--stars to the Eddington limit 
(Lamers and Leitherer \cite{la93}; Herrero et al. \cite{He99}) and of the large effect
of metal line blanketing not usually accounted for in the
atmosphere models of massive stars (Lanz et al. \cite{la96}), some discrepancy
seems to be still present.
The remaining mass discrepancy may arise, in part, 
from the use
of non--rotating models for deter\-mining the evolutionary masses. High rotation
produces larger He--cores and He--rich envelopes, all that
implies overluminous stars with lower gravity as evolution proceeds.
This can occur even
for the slow rotating objects because they could have had a sufficiently high initial rotational velocity.
The observation that the mass discrepancies are 
found only for low gravity objects may
reflect the fact that rotation implies more and more important changes in the $\log g_{\rm eff}$ versus $\log T_{\rm eff}$
plane when evolution proceeds.
Indeed, looking at Fig.~\ref{fgeff} below one sees that at high gravity (i.e. at an early evolutionary stage),
one does not expect any mass difference when using rotating or non--rotating tracks. In contrast,
the mass difference between the rota\-ting and the non--rotating tracks become more and more
important in the high mass star range and for the low values of $\log g_{\rm eff}$.
The rotating 40 M$_\odot$ track crosses the non--rotating 60 M$_\odot$ model on Fig.~\ref{fgeff}
indicating that differences of $\sim$30\% in mass are quite possible.

For the fast rotating objects, a part of the mass discrepancy may also 
be due to a possible underestimate of the gravity (see Sect. 7.1). Indeed,
an underestimate of the gravity would also imply an underestimate of the
spectroscopically determined masses (Herrero et al. 2000).

\subsection{Lifetimes and isochrones}

Table~\ref{tbl-1} presents some properties of the models. Column 1 and 2 give the initial mass and the initial velocity $v_{\rm ini}$ respectively.
The mean equatorial rotational velocity $\overline{v}$ during the MS phase is indicated in column 3.
This quantity is defined by
 
$$\overline{v}=1/t_{H} \int_{0}^{t_{H}} v(t) dt,$$ 

\noindent where $t_{H}$ is the duration of the H--burning phase given in column 4, except for those stars which enter the Wolf--Rayet
phase while still burning their hydrogen in their core, i.e. for the rotating 60 and 120 M$_\odot$ models, for which
$t_{H}$ has been replaced by the duration of the O--type star phase.
The
H--burning lifetimes $t_H$, the masses M, the equatorial velocities $v$, the helium surface abundance $Y_s$ and the 
surface ratios (in mass) N/C and N/O at the end of the H--burning phase are given in columns 4 to 9.
The columns 10 to 13 present some properties of the models when the star is a blue supergiant (BSG) or an LBV star.
For stellar models with M$\le 40$ M$_\odot$, the ``BSG stage'' in Table~\ref{tbl-1} corresponds either to the stage when
$\log T_{\rm eff}=4.0$ during the first crossing of the Hertzsprung--Russel diagram or to the bluest point
on the blue loop if any for M$\le 12$ M$_\odot$. For non--rotating stellar models with M $\ge 60$ M$_\odot$, the ``LBV stage''
corresponds to the point when the star has lost half of the matter ejected by the stellar winds 
between the end of the MS and the entrance into the W--R phase. In the case of rotating models,
the ``LBV stage'' corresponds to the period during which the surface velocity becomes critical
and huge mass loss rates ensue. Again, here we choose a model in the middle of this phase.
The columns 14 to 19 present some characteristics of the stellar models at the end of the He--burning phase and
$t_{He}$ is the He--burning lifetime.
In Table~\ref{tbl-2} some properties of 20 M$_\odot$ models at the end of the MS
are indicated, $v_{\rm ini}$ and $\overline{v}$ have the same meaning as above.

\begin{table}
\caption{Properties of 20 M$_\odot$ models at the end of the MS for different initial
velocities. The velocities are in km s$^{-1}$, the lifetimes in million years,
the masses in solar mass and the abundances in mass fraction.} \label{tbl-2}
\begin{center}\scriptsize
\begin{tabular}{ccrccccc}
$v_{\rm ini}$ & $\overline{v}$ & $t_H$ & M & $v$ & Y$_s$ & N/C & N/O \\
      &      &         &         &       &       &       &      \\
\hline
      &      &         &         &       &       &       &      \\
 0 &       0    & 7.350  & 19.019 & 0     & 0.30  & 0.25  & 0.12 \\
 50 &      30    & 7.720  & 18.896 & 18    & 0.30  & 0.27  & 0.12 \\
 100 &      62    & 8.292  & 18.681 & 46    & 0.30  & 0.45  & 0.19 \\
 200 &     132    & 8.901  & 18.324 & 94    & 0.32  & 1.01  & 0.38 \\
 300 &     197    & 9.309  & 18.020 & 167   & 0.35  & 1.77  & 0.58 \\
 400 &     253    & 9.745  & 17.646 & 217   & 0.37  & 2.54  & 0.76 \\
 500 &     294    & 10.275 & 17.181 & 213   & 0.40  & 3.65  & 0.99 \\
 580 &     304    & 10.324 & 17.148 & 214   & 0.39  & 3.75  & 1.00 \\
     &       &         &         &       &       &       &      \\

\hline
\end{tabular}
\end{center}

\end{table}

From Table~\ref{tbl-1} one sees that for $Z=0.020$ the  lifetimes are 
increased by about 20--30\% when the mean rotational velocity 
on the MS increases from 0 to $\sim$200 km s$^{-1}$. This
modest increase is explained by the fact that
even if there is more fuel available in the core,
the luminosity is also increased.
From the data presented in Table~\ref{tbl-2}, one can deduce a nearly
linear relation between the relative enhancement of the MS lifetime,
$\Delta t_{\rm H}$ and $\overline{v}$, where 

$$\Delta t_{\rm H}(\overline{v})=[t_{\rm H} (\overline{v})-t_{\rm H}(0)]/t_{\rm H}(0).$$
 
\noindent One obtains,

$${\Delta t_{\rm H} (\overline{v}) \over t_{\rm H}(0)}=0.0013\cdot \overline{v},$$

\noindent where $\overline{v}$ is in km s$^{-1}$. This relation reproduces 
the values of $\Delta t_{H} (\overline{v})/t_{H}(0)$ from table~\ref{tbl-2} with
an accuracy better than 5\%. It also applies with the same accuracy
to the values listed in Table~\ref{tbl-1} for the masses between 15 and 40 M$_\odot$. 
The He--burning lifetimes are less affected by rotation than the MS lifetimes. 
The changes are less than 10\%. The ratios $t_{He}/t_{H}$ of the
He to H--burning lifetimes are only slightly decreased by rotation and
remain around 10--15\%.

\begin{figure}[tb]
  \resizebox{\hsize}{!}{\includegraphics{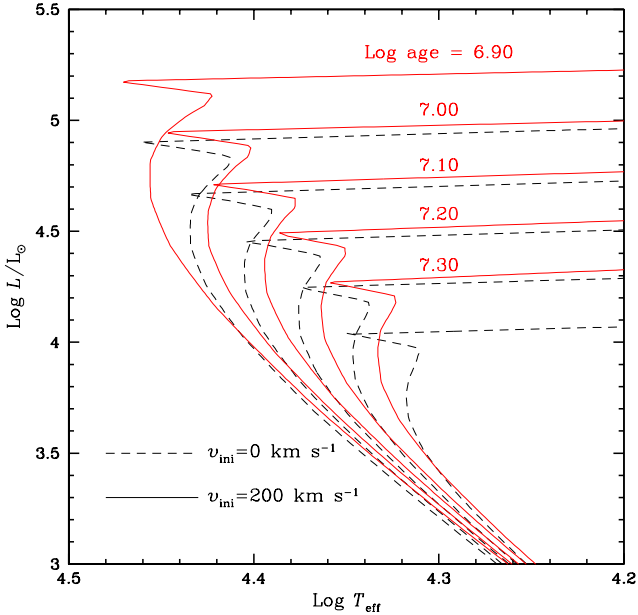}}
  \caption{Isochrones computed from stellar evolutionary tracks for the solar metallicity. 
The dashed and continuous lines correspond to the case of non--rotating and rotating 
stellar models respectively. In this last case, models have
an initial velocity $v_{\rm ini}$ of 200 km s$^{-1}$. The logarithms
of the ages (in years) label the isochrones computed from the models with rotation.}
  \label{isoc}
\end{figure}

On Fig.~\ref{isoc} isochrones for ages between about 8 and 20 10$^6$ yr, 
computed from non--rotating and rotating stellar models are presented. The ``rotating'' isochrones are
computed from the models with an initial rotational 
velocity $v_{\rm ini}$ of 200 km s$^{-1}$ on the ZAMS.

At a given age, the upper part of the ``rotating'' isochrones are bluer
and more luminous. Typically, for an age equal to about $20 \cdot 10^6$ years  ($\log$ age =7.3) 
the reddest point on the MS is shifted by 0.015 dex in $\log T_{\rm eff}$ and by - 0.6 
in $M_{\rm bol}$ when rotation is taken into account. 
An isochrone with rotation is almost identical to an isochrone without rotation with log age smaller
by 0.1 dex.
This has for consequence that
rotation slightly increases the age associated to a given 
cluster. From Fig. 9, one sees that the ``rotating'' isochrone 
for an age equal to $20 \cdot 10^6$ years has the same luminosity at the turn--off 
than the ``non-rotating'' isochrone for an age equal to $16 \cdot 10^6$ years. Thus rotation
increases the age estimate by about 25\%. We may wonder
whether this effect explains the age difference between the estimates based on the upper
MS and the estimates based on the lithium content of the very low mass stars
(see e.g. Martin et al. 1998; Barrado y Navascu\'es et al. 1999).
One must  also account for the dispersion of rotational velocities
and possibly of the orientation angles.
These two effects introduce some dispersion in the way stars 
are distributed in the HR diagram and thus affect the 
interpretation of the clusters' observed sequences (cf. Maeder \cite{ma71}). 

If a bluewards track occurs, as for very massive stars with fast rotation,
the larger core and mixing lead to
much longer lifetimes in the H--burning phase.
In this case, the fitting of time--lines becomes hazardous.

\section{Evolution of the rotational velocities} 

\subsection{Model results for stars with a large mass loss}

\begin{figure}[tb]
  \resizebox{\hsize}{!}{\includegraphics{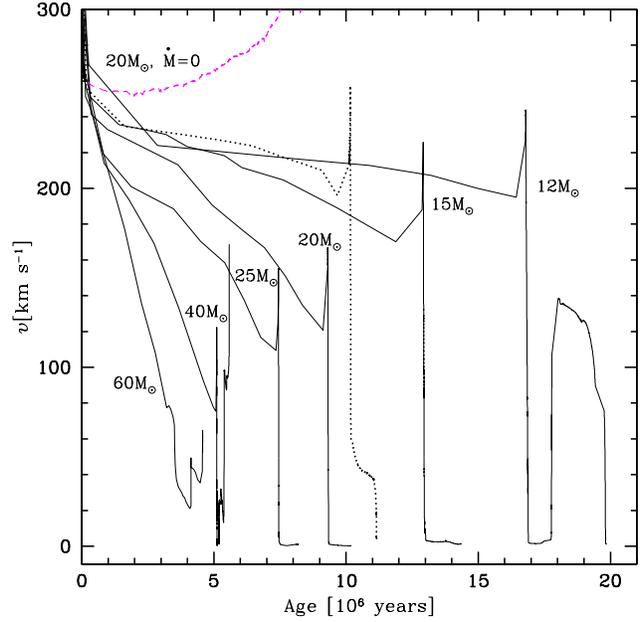}}
  \caption{Evolution of the surface equatorial velocity as 
a function of time for stars of different initial masses
with $v_{\rm ini}$ = 300 km s$^{-1}$. 
The continuous lines refer to solar metallicity models, 
the dotted line corresponds
to a 20 M$_\odot$ star with $Z$ = 0.004.   
The dashed line  corresponds to a 20 M$_\odot$ star
 without mass loss.
}
  \label{v/age}
\end{figure}

\begin{figure}[tb]
  \resizebox{\hsize}{!}{\includegraphics{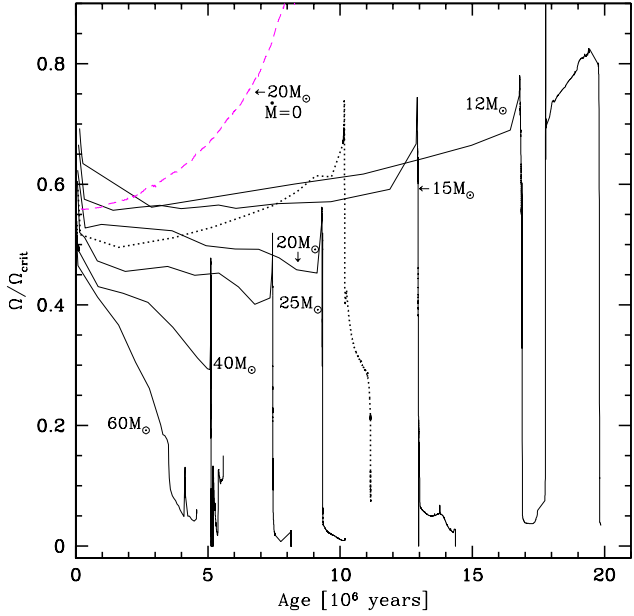}}
  \caption{Same as Fig.~\ref{v/age} for the ratio $\Omega/\Omega_{\rm crit}$ 
of the angular velocity to the break--up velocity
at the stellar surface.
}
  \label{OM/age}
\end{figure}

\begin{figure}[tb]
  \resizebox{\hsize}{!}{\includegraphics{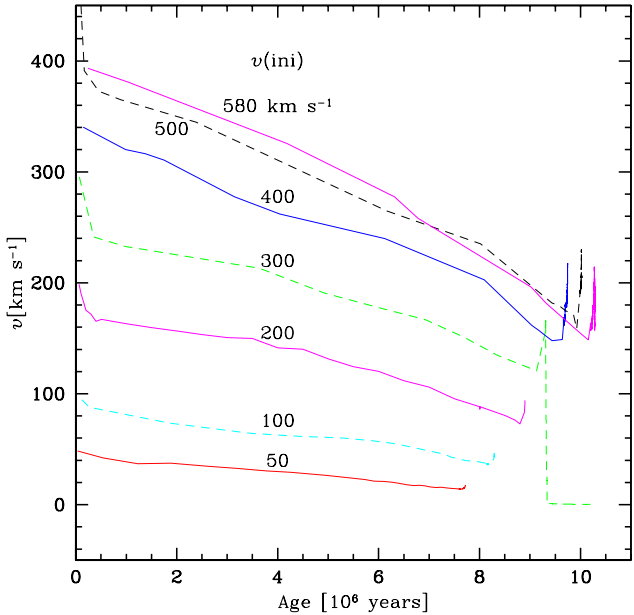}}
  \caption{Evolution of the surface equatorial velocity as 
a function of time for 20 M$_\odot$ stars with different
initial velocities. 
}
  \label{v20/age}
\end{figure}

Figs.~\ref{v/age} and \ref{OM/age} show the evolution of $v$
and of $\frac{\Omega}{\Omega_{\rm{crit}}}$ as a function of the age for 
the present models. Firstly,
we notice that for models without mass loss, as shown for the
20 M$_{\odot}$ with $\dot{M}$ = 0,  $v$
and $\frac{\Omega}{\Omega_{\rm{crit}}}$ go up fastly 
so that the critical velocity
would be reached near the end of the MS phase. The current model of 
20 M$_{\odot}$ with mass loss show a significant decrease of $v$,
while the critical ratio remains almost constant during most of the MS phase.
 Figs.~\ref{v/age} and \ref{OM/age} show how fastly rotation decreases 
at the surface of the most massive stars, which lose a lot of mass. 
Consistently we see that the reduction of the surface rotation is 
much larger for the more massive stars. This is of course a 
consequence of the removal of large amounts of angular momentum by the
stellar winds. The effect is amplified by the increase of the mass loss
in fast rotators (Eq.~2). We see that the decrease of $\frac{\Omega}{\Omega_{\rm{crit}}}$
is so strong that
it will  prevent a massive star to reach the critical velocity
near the end of the MS phase.  If the star makes
extended excursions in the HR diagram at the end of the 
MS like is the case for the 60 M$_{\odot}$ model,
 then it may reach the critical velocity. The specific case of 
stars close to the $\Omega$--limit will be examined in a future study,
since this requires some further theoretical developments.

Let us mention here that the present results differ from those
obtained by Sackmann and Anand (\cite{Sa70}) and
Langer (\cite{La97}, \cite{La98}). Indeed, these authors find that 
the star reaches the 
break--up limit during the MS  phase. As an example, in a 60 M$_{\odot}$
star, even a model with an initial $v_{\rm{ini}}$ of 100 km s$^{-1}$
reaches the break--up limit near the end of the MS--phase (Langer 1998). 
This result
leads Langer to conclude that most massive stars may reach the break--up
limit or the so--called $\Omega$--limit during their MS evolution.
Let us however emphasize here that such a conclusion is based 
on a particular
definition of $v_{\rm crit}$ still subject to discussion (see Sect. 2.4),
on the assumption of solid body rotation,
and on models not accounting for the effects of rotationnally induced mixing.
We see here that modifying these hypothesis (and also
using other prescriptions for the mass loss rates) lead to very different results. 
One of the first step to clarify the situation 
is to determine
which expression for $v_{\rm crit}$ (cf. Glatzel 1998; Langer 1998)
is the correct one and how rotation
affects the mass loss rates. These developments, now in progress, will be
particularly needed for the study of the evolution of the most massive stars, like a 120 M$_\odot$ model, which have a high value of the Eddington factor.
Also this is important for the formation and evolution of W--R stars, which will
be studied in a further work.

\begin{figure*}[tb]
  \resizebox{\hsize}{!}{\includegraphics[angle=-90]{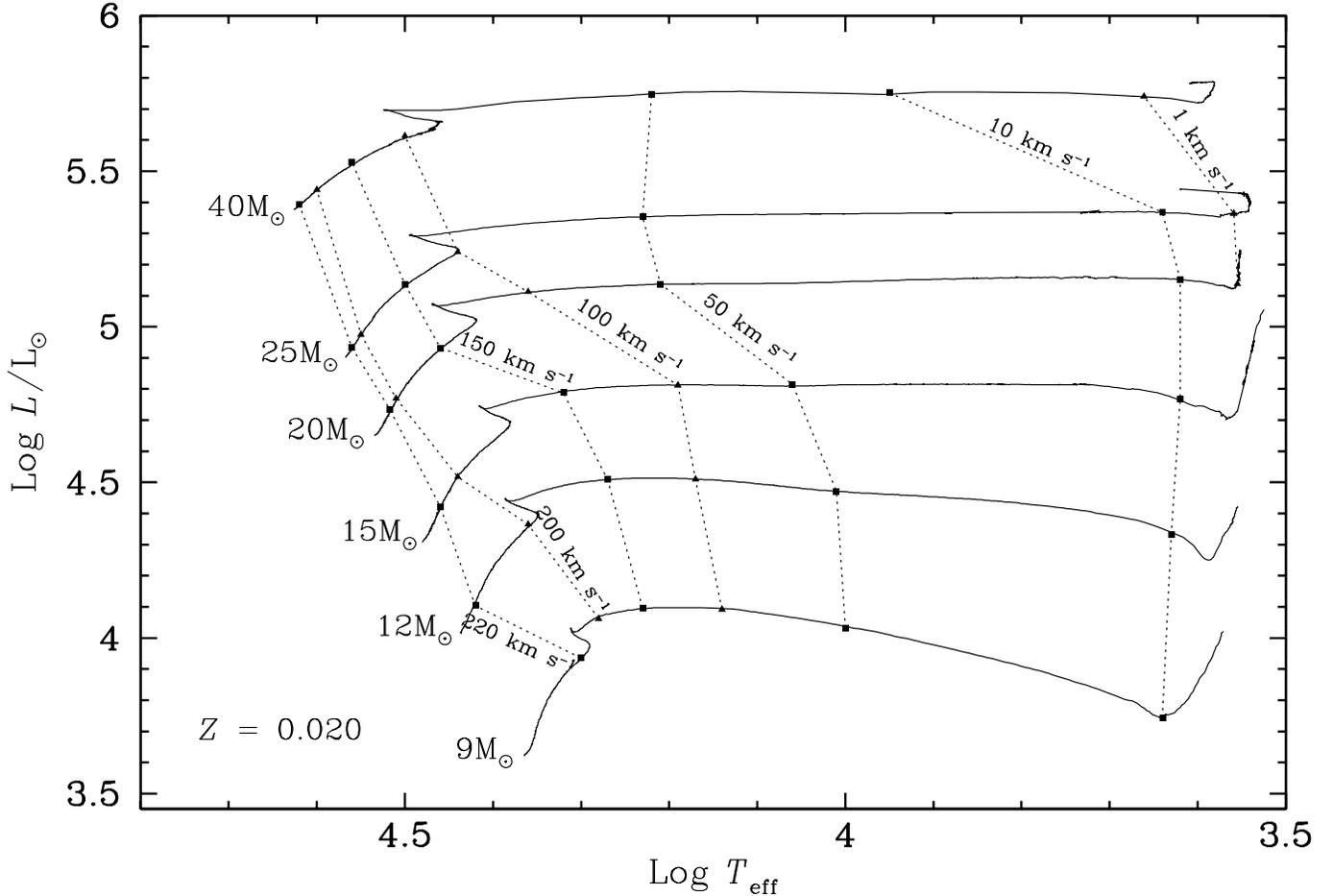}}
  \caption{Evolution of the equatorial surface velocities along the evolutionary
tracks in the HR diagram starting from $v_{\rm ini}$ = 300 km s$^{-1}$.
For purpose of clarity, only the first part of the 40 M$_\odot$ track
is shown.
}
  \label{v_HR}
\end{figure*}

Fig.~\ref{v20/age} shows the evolution of 
$v$ with age for models of a 20 M$_{\odot}$ star
with different initial velocities $v_{\rm ini}$ from
50 to 580 km s$^{-1}$. We see that the decrease in
the surface $v$ is larger for larger initial rotation.
This is a consequence of the larger mass loss rates in fast rotators.
We notice some convergence (cf. Langer 1998) of the curves for the large
initial  $v_{\rm{ini}}$. This convergence would be more
pronounced and affect the models with lower $v_{\rm ini}$ if
the dependence of $\dot M$ vs $v$ would be stronger than given
by Eq.~(2).
This certainly reduces the scatter of 
$v$ near the end of the MS phase, but does not produce a full
convergence. 

\subsection{Results for stars with lower mass loss}

The models of 12 and 15 M$_{\odot}$ show curves in  
Figs.~\ref{v/age} and \ref{OM/age} with little reduction of
 $v$, while there are slight increases of the critical ratio
 $\frac{\Omega}{\Omega_{\rm{crit}}}$ during MS evolution. A peak is reached
during the overall contraction phase at the end of the MS phase. Then, 
$v$ and the critical ratio go down as the star moves
to the   red supergiant phase.
 During such a fast evolution, we may say that the
rotation evolves almost like the case of simplified models,
where the angular momentum is conserved locally.
The large growth of the radius just implies a decrease of 
$v$ and of $\frac{\Omega}{\Omega_{\rm{crit}}}$.  Later
during the blue loops, where the Cepheid instability strip is crossed, 
the rotation
velocity becomes very large again and could easily become close 
to critical. This behaviour, also found by Heger \& Langer (1998),
results from the stellar contraction 
which concentrates
a large fraction of the angular momentum of the star
 (previously contained in the
extended convective envelope of the RSG) 
in the outer few hundredths of a solar mass. 
This result suggests that rotation may also somehow influence the
Cepheid properties, in addition to the increase of stellar
luminosity discussed above.

Amazingly, we notice that the stars with initially low mass loss
during the MS phase, like for stars with M $\leq 12$ M$_{\odot}$,
have more chance to reach the  break--up velocities and thus huge mass loss
than the more massive stars which lose a lot of mass on the MS. 
It is somehow surprising that little mass loss during the 
MS may favour large mass loss rates at the end of the 
MS phase. This may explain why 
 stars close to break--up, like the Be stars,  do not form among
the O--type stars, but  mainly among the B--type stars, where we see that 
the ratio $\frac{\Omega}{\Omega_{\rm{crit}}}$ may increase during 
the MS phase. Another related observation is the fact that
the relative number of Be--stars
with respect to B--type stars is much higher in the LMC and SMC than
in the Galaxy (cf. Maeder et al. \cite{MGM99}). In the LMC and SMC,
 due to the lower metallicity, the average  mass loss rates 
are lower and thus these stars may 
 keep higher rotation in general and thus form more Be stars.
 This explanation  does not exclude differences
in the distribution of $v_{\rm{ini}}$ as well.

\subsection{The rotational velocities in the HR diagram}

The evolution of $v$ in the HR diagram is shown in
Fig.~\ref{v_HR}. Starting with models having a velocity of 300 km s$^{-1}$ on the
zero--age main sequence, we give some  lines of constant $v$
over the HR diagram. On the MS, we notice 
 in particular that the decrease is
much faster for the most massive stars than for stars with
M $\leq$ 15 M$_{\odot}$. 
This difference remains also present in the domain of B--supergiants.
During the crossing of the HR diagram, the rotational velocities decrease
fastly, to become very small, i.e. of the order of a few km s$^{-1}$, in the
red supergiant phase. 

It is beyond the
scope of this paper to make detailed comparisons of the evolution 
of the distribution of the velocities over the HR diagram,
however, we may notice a few points. The fact that the average 
$\overline v$ is lower for O--type stars than for the early
B--type stars (Slettebak, \cite{Sl70}) may be the consequence of the higher 
losses of mass and angular momentum in the most massive stars.
Also, we remark that the increase of $\overline v$ from O--stars to
B--stars is larger for the stars of luminosity class IV than for class V 
(Fukuda, \cite{Fu82}). This is consistent with our models,
which show (cf. Fig.~\ref{v_HR}) that
 the differences  of $\overline v$ beween O-- and 
B--type stars are much larger at the end
of the MS phase.  Another fact in the observed data is the
 strong decrease
of $\overline v$ for the massive supergiants of OB--types. 
This is predicted by all stellar models (cf. also Langer \cite{La98})
due to the growth of the stellar radii. 
Further detailed  comparisons may  perhaps provide some new tests and constraints.

\section{Evolution of the surface abundances}

\begin{figure*}[tb]
  \resizebox{\hsize}{!}{\includegraphics[angle=-90]{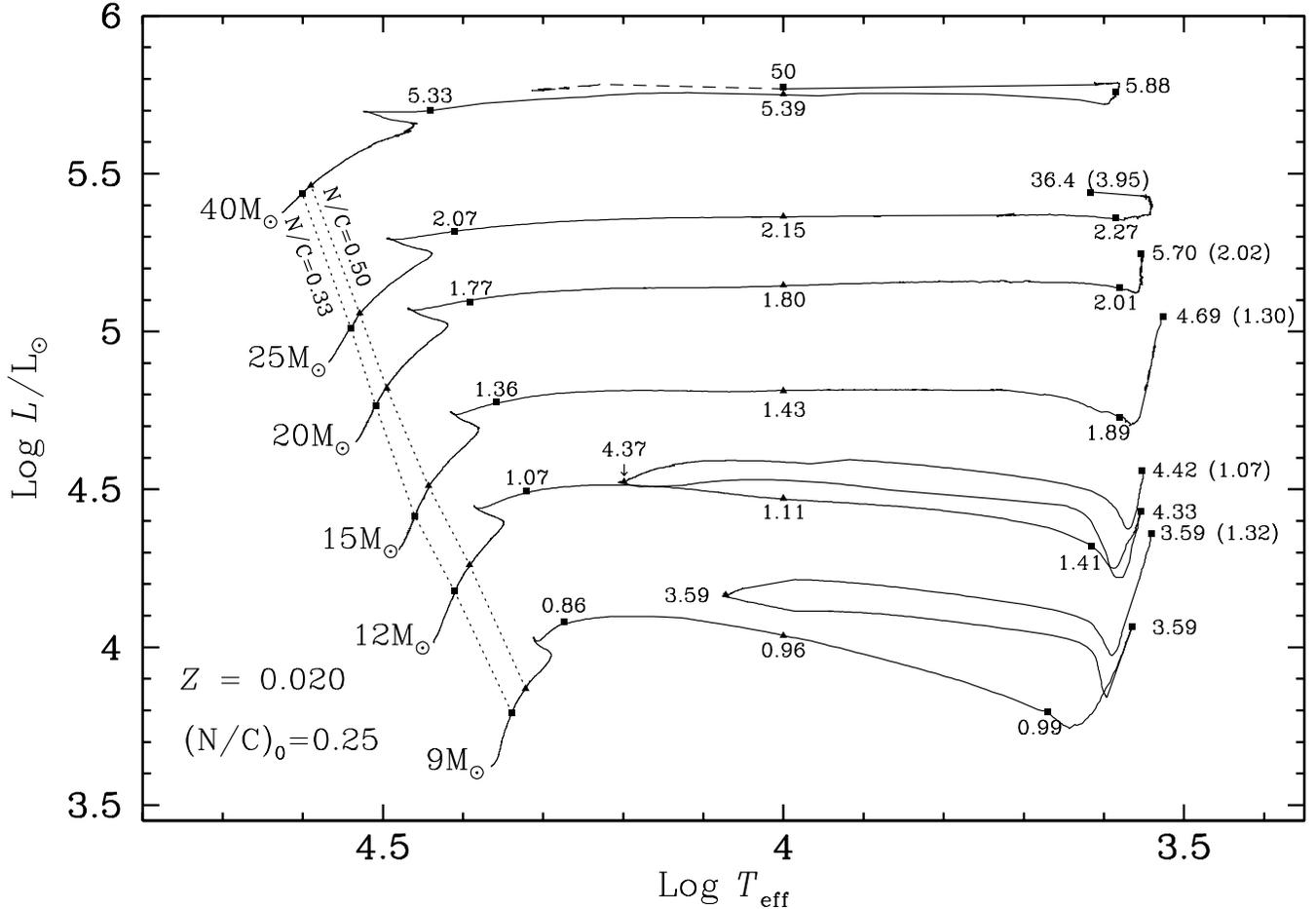}}
  \caption{Evolution of the N/C ratios
(in mass) at the surface of rotating models for $v_{\rm ini}$ = 300 km s$^{-1}$.
Dotted lines joining evolutionary stages with the same value of
the N/C ratio are superposed on the MS evolutionary tracks. 
Values of
the N/C ratios are indicated at some points during the post--MS phases. The numbers indicated
in parentheses correspond to the values of the N/C ratio at the surface of the non--rotating models
at the end of the He--burning phase. The initial ratio (N/C)$_0$ is equal to 0.25.
For purpose of clarity, only the first part of the 40 M$_\odot$ track
is shown.
}
  \label{HRcn}
\end{figure*}

The chemical abundances offer a very powerful test of internal
evolution and they give strong evidences in favour of
some additional mixing processes in massive stars.
A review of the observations may be found in Maeder and Meynet (2000). Here
we shall concentrate on the discussion of the theoretical results and we shall compare
them with some recent observations.

The most striking feature appearing in Tables~\ref{tbl-1} and \ref{tbl-2} as well as on
Fig.~\ref{HRcn} 
is the change of surface
abundances in rotating stellar models
(cf. Langer 1992). The He--, N--enrichments and the related C-- and O--depletions 
at the surface already occur on the MS. 
The more massive the star is, the more pronounced
are the enrichments, supporting the expectation that mixing becomes more and
more efficient when the mass increases (Maeder \cite{ma98}). The same is true when the
initial rotational velocity increases.
During the crossing of the Hertzsprung--Russel diagram, the evolution is sufficiently short
for not allowing any important change of the surface abundances (see Fig.~\ref{HRcn}). Further changes occur when the star
becomes a red supergiant and undergoes the first dredge--up. 
In contrast,
non--rotating stellar models show no change of the surface abundances until the first dredge--up 
in the red supergiant stage (see Table~\ref{tbl-1}). 
This implies that these models predict
some enrichment neither during the MS nor in the blue supergiant phase unless a blue loop is formed.
Moreover, as can be seen in Fig.~\ref{HRcn}, the ratios obtained at the end of the He--burning phase
are significantly lower than the ones obtained in rotating models.

\subsection{He--enrichments and $v\sin i$}

\begin{figure}[tb]
  \resizebox{\hsize}{!}{\includegraphics{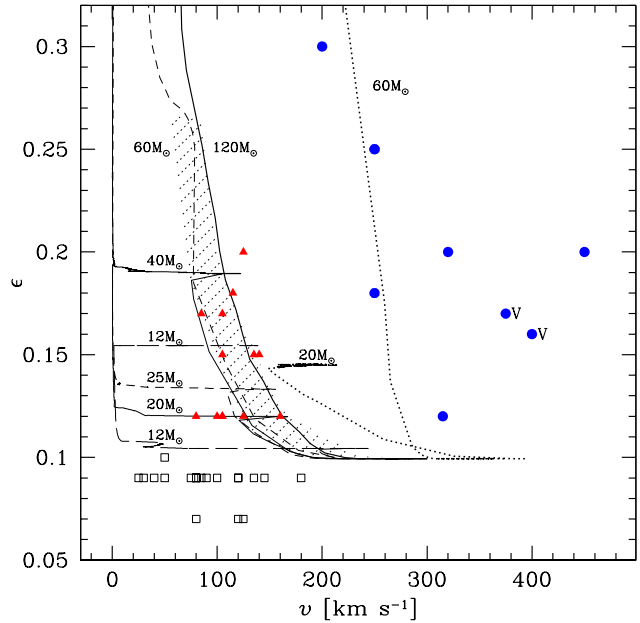}}
  \caption{Evolution of the helium surface abundance as a function of the equatorial
rotational velocity; $\epsilon = {n({\rm H_e}) \over n({\rm H})+n({\rm He})}$ where $n({\rm H_e})$ and $n({\rm H})$ are
the abundances in number of helium and hydrogen. All the tracks except two were
computed for an initial velocity of $v_{\rm ini}$ = 300 km s$^{-1}$. The dotted tracks at the right
of the figure correspond to fast rotating models, namely a 20 and a 60 M$_\odot$ model with $v_{\rm ini}$ equal to
580 and $\sim$400 km s$^{-1}$ respectively. Only the MS of the fast rotating 20 M$_\odot$ model is shown.
The observed points ($\epsilon, v\sin i$)
are from Herrero et al. (\cite{He92}, \cite{He99}, \cite{He20}). The empty squares are for stars with no (or very low) 
He--enrichment at the surface, the solid triangles are for objects whose surface present He--enrichments and
having $v\sin i <$ 200 km s$^{-1}$, the solid circles correspond to He--enriched stars with
$v\sin i \ge$ 200 km s$^{-1}$. A V labels the stars of this last group classified as MS stars.
}
  \label{ysvit}
\end{figure}

Fig.~\ref{ysvit} shows the evolution in the $\epsilon$ versus 
surface velocity plane where $\epsilon={n(\rm{H_e}) \over n({\rm H})+n({\rm He})}$, 
$n({\rm H_e})$ and $n({\rm H})$ being
the abundances in number of helium and hydrogen at the surface.
The theoretical tracks correspond to the equatorial
velocities $v$, while the observed points
from Herrero et al. (\cite{He92}, \cite{He99}, \cite{He20}) are $v\sin i$.
This implies that the observed values are smaller
on the average by a factor $\pi/4$ with respect to the theoretical values.

In Fig.~15, the tracks go generally from the bottom on the right to the
top on the left. Along a given track, the He--enrichment increases when the velocity decreases.
The shaded zone in Fig.~\ref{ysvit} corresponds to the MS 
for the models with $v_{\rm ini}$ = 300 km s$^{-1}$.  
The higher initial mass, the greater the He--enrichments which can be
reached during the MS. We see also that during the MS phase, for initial masses
inferior to about 60 M$_\odot$, all the tracks with $v_{\rm ini}$ = 300 km s$^{-1}$
follow more or less the same $\epsilon$ versus $v$ relation.
However, the relation is changed when the initial velocity
is different (see the fast rotating 20 and 60 M$_\odot$ models).
The surface He--enrichments on the MS generally depend
on the following factors: the initial mass, the initial metallicity 
(Maeder \& Meynet 2000b; Meynet 2000a), the initial velocity
and the age of the star.

A low surface velocity at a given evolutionary stage
does not exclude that
in the past the star was a fast rotator. The slow rotation may result
from the loss of angular momentum by stellar winds and/or from the increase
of radius of the star. On the other hand, a high velocity in the past implies 
He-- and N--enrichments of the surface. 

\begin{figure}[tb]
  \resizebox{\hsize}{!}{\includegraphics{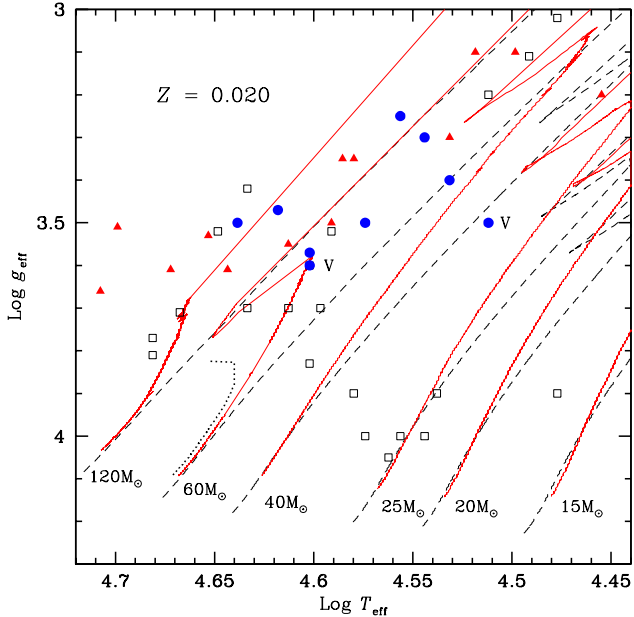}}
  \caption{Evolutionary tracks in the $\log g_{\rm eff}$ versus $\log T_{\rm eff}$ plane where $g_{\rm eff}$ is the effective surface
gravity. The dashed and continuous lines are for the non--rotating and rotating ($v_{\rm ini}$ = 300 km s$^{-1}$)
models respectively. The dotted track corresponds to the fast rotating 60 M$_\odot$ plotted on Fig.~\ref{ysvit}.
Only the beginning of the evolution is shown.
The positions of the stars observed by Herrero et al. (\cite{He92}, \cite{He99}, \cite{He20}) are
indicated with the same symbols as in Fig.~\ref{ysvit}.
}
  \label{fgeff}
\end{figure}

Let us compare in Fig.~\ref{ysvit} the theoretical tracks in the $\epsilon$ versus $v$ plane
with the observations of OB stars performed by Herrero et al. (\cite{He92}, \cite{He99}, \cite{He20}).
Recent works (McErlean et al. \cite{mc98}; Smith and Howarth \cite{sm98}) indicate that
accounting for a microturbulent velocity line broadening in the model atmosphere reduces
the derived He--abundances for supergiants later than O9. However, according to Villamariz and Herrero (\cite{vi99}) this
effect cannot explain all the observed overabundances, especially for the earlier types.
On Fig.~\ref{fgeff} the observed points are plotted in the $\log g_{\rm eff}$ versus $\log T_{\rm eff}$
plane where $g_{\rm eff}$ is the effective surface gravity. We estimate
$g_{\rm eff}=\frac {GM} {R^2}-\Omega^2 R \sin \theta $ for the average
orientation angle. 
Let us note that for
the models plotted in Fig.~\ref{fgeff}, there is little difference
between the effective gravities at the pole and at the equator. Indeed
the ratio between these two gravities never exceeds 1.3 which means a 
vertical dispersion of about 0.1 dex in Fig.~16.
Non--rotating and rotating evolutionary tracks
are superposed to the observed points in Fig.~\ref{fgeff}. Since
most of the enriched stars are in the vicinity of the 
120 and 60 M$_\odot$ tracks (see Fig.~\ref{fgeff}), we can
wonder wether the changes of the surface abundances can be explained as an effect of mass loss only.
It does not seem to be the case, because 
the part of the track shown on Fig.~\ref{fgeff} for the  non--rotating 60 M$_\odot$ 
model presents no surface He--enrichment. In the case of the non--rotating 120 M$_\odot$ model
only the part of the track with $\log g_{\rm eff}$ inferior to 3.3 has $\epsilon > 0.14$.
Thus the He--enrichments cannot be accounted for by current evolutionary models
as was already pointed out by Herrero et al. (\cite{He92}, cf. also Maeder 1987).
In the following we shall
suppose that these enhancements are due to rotation.

Let us consider four groups of stars. In the first group
we place all the stars presenting no He--enrichment at their surface ($\epsilon \le 0.12$)
and having $v \sin i  < 200$ km s$^{-1}$ 
(empty squares on Figs.~~\ref{ysvit} and \ref{fgeff}), 
in the second one are the enriched
stars ($\epsilon >$ 0.12, solid triangles) having $v \sin i  < 200$ km s$^{-1}$. The third group
(which is empty at present) consists of the fast rotators with no He--enrichment.
These stars would occupy the bottom right corner in Fig.~\ref{ysvit}. Finally, 
the fourth group contains the He--enriched stars
with $v \sin i \ge 200$ km s$^{-1}$ (solid circles). 
\vskip 2mm
\noindent {\bf Group 1}
\vskip 2mm
The non--enriched stars with low rotation
can be interpreted either as stars with small initial
velocities or as young
fast rotators whose surface has not yet been enriched in helium by rotational mixing
(see also Herrero et al. \cite{He99}). 
In this respect let us mention that all the stars having $\log g_{\rm eff} > 3.7$, and which therefore
are probably not too evolved, present no He--enrichment.
\vskip 2mm
\noindent {\bf Group 2}
\vskip 2mm
The most striking feature of the second group of stars ($\epsilon > 0.12$ and
$v \sin i < 200$ km s$^{-1}$) is the fact that they are distributed in a relatively narrow
range of $v \sin i$ between 80 and 160 km s$^{-1}$. This may result from 
the following facts: firstly there exists a minimum value of the initial velocity 
for rotational mixing to be able to drive changes of the helium surface 
abundances during the H--burning phase. 
Secondly the observed distribution also reflects the way the surface velocity declines during
the H--burning phase and the narrow range of observed velocities may result from
some convergence effect as the one mentionned in Sect. 6.1 (see also Fig.~12).
We see that the high $\epsilon$ values reached by some
of these stars (superior to $\sim$0.16) would be compatible with their high initial mass
implied from their position in Fig.~\ref{fgeff}.  
\vskip 2mm
\noindent {\bf Group 3}
\vskip 2mm
Very interestingly no stars are observed with a $v\sin i \ge 200$ km s$^{-1}$ and no
He--enrichment. Keeping in mind that the observed stars do not represent a statistical complete sample,
one can nevertheless wonder why no stars are observed in this zone. A possibility 
would be that these very fast rotators do not exist, but this is not realistic.
Indeed stars with velocities as high as 300 km s$^{-1}$ and more have been observed on the MS
(e.g. Penny \cite{pe96}; Howarth et al. \cite{ho97}; see also Fig.~\ref{ysvit}). Moreover if such stars were not formed
how to explain the stars in Group 4, having a very high $v\sin i$ and
an important He--enrichment (the solid circles in Fig.~\ref{ysvit}) ? 
These stars are likely the descendants of
very fast MS rotating stars (see the discussion in the subsection Group 4 below). 

The fact that no stars are observed in the Group 3 
may indicate that the change of the surface abundances occur within
a small fraction of the visible MS life. 
For the fast rotating
60 M$_\odot$ model plotted on Fig.~\ref{ysvit}, the surface retains its
initial composition ($\epsilon < 0.11$) only during a third of its H--burning lifetime. More rapidly rotating models
would still reduce the fraction of the MS time spent with no change of the surface abundances.
The lack of fast rotators with normal surface composition may be due to the fact that
young massive fast rotators are still embedded
in the cloud from which they formed. The models of massive star formation with accretion (Bernasconi
and Maeder \cite{be96}) indicate
that when the stars become visible, they have already burnt some fraction of their central hydrogen,
thus some transport of He and N to the stellar surface may have occured.

\vskip 2mm
\noindent {\bf Group 4} 
\vskip 2mm
Could the stars belonging to the fourth group
($\epsilon > 0.12$ and $v\sin i \ge 200$ km/ s)
be formed by stars previously in the low velocity range and which have been
accelerated for instance by contraction on a blue loop~?
The answer is likely no. Indeed
blue loops cannot accelerate the surface beyond the critical 
velocity which, for a 12 M$_\odot$ blue supergiant at the
tip of a blue loop,
is of the order of 250 km s$^{-1}$. In addition some stars 
in the group 4 are classified as MS stars (see the stars
labeled with a V on Figs.~\ref{ysvit} and \ref{fgeff}).
Another possibility would be to consider the stars in Group 4 as
secondary
stars in close binary systems which would have been accelerated
through the process of mass accretion ? But these stars are not observed to
belong to binary systems. Therefore
the most reasonable hypothesis is to consider the stars in Group 4 as the
natural descendants of
very fast MS rotating stars.

Their chemical enrichment at the surface is very fast.
Their chemical structure
as a result of the strong rotational mixing is probably near homogeneity.
This view is supported by the fact that high values of $v\sin i$
 are observed for very high values of $\epsilon$
implying that the surface velocity does not decrease too much 
in the course of the evolution. This can be accounted for if the star remains compact,
{\it i.e.} in the blue part of the HR diagram as is the case for a strongly mixed star  
(Maeder \cite{ma87}; see also the fast rotating 60 M$_\odot$ track in Fig.~\ref{ysvit}). 
Another effect could also be important in that respect, {\it i.e.} the anisotropy 
of the stellar winds when stars are rotating near break--up. For the hot stars, the von Zeipel (\cite{vZ24})
theorem implies that most of the mass is ejected from the pole (Maeder \cite{Mae99}). This prevents
the loss of important angular momentum and maintains a high surface velocity.

If the stars in the fourth group are nearly homogeneous objects, one would expect higher 
effective gravities than observed. Typically,
the fast rotating well mixed 60 M$_\odot$ model remains at a high value of $\log g_{\rm eff}$, 
at least for the portion of the evolution computed here (see Fig.~\ref{fgeff}).
Moreover, as noted above, the maximum dispersion in $\log g_{\rm eff}$ due to orientation effects is around 0.1 dex.
Could the $\log g_{\rm eff}$ be
underestimated ? Even if it is difficult on the base of the present data
to ascertain such a point of view, one can
note that the $\sin i$ for these fast rotating stars cannot
be too far from 1, otherwise one would obtain surface velocities above the break--up value.
This means that these stars are seen essentially equator--on. The $\log g$ in the equator band
is inferior to the surface averaged $\log g$ and thus the observed $\log g$ might
be underestimated. 
Of course a quantitative analysis requires a detailed study
of how the variation of $\log g$ with the latitude affects the spectroscopically determined
gravities. 

\begin{figure}[tb]
  \resizebox{\hsize}{!}{\includegraphics{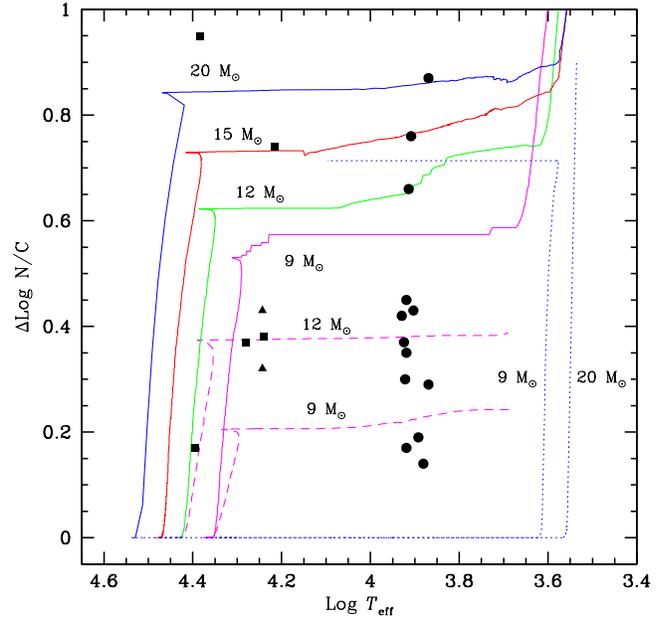}}
  \caption{Evolution as a function of log $T_{\rm eff}$ of
$\Delta \log {\rm N\over \rm C}= \log ({\rm N/C})-\log ({\rm N/C})_i$, where N and C
are the surface abundances (in number) of nitrogen and carbon
respectively, the index $i$ indicates initial values. The continuous tracks
correspond to rotating stellar models with $v_{\rm ini}$ = 300 km s$^{-1}$. The
dashed and dotted lines are respectively for initial $v_{\rm ini}$ = 200 and
0 km s$^{-1}$. Only parts of the tracks with $v_{\rm ini}$ = 200 km s$^{-1}$ are shown.
The initial masses are indicated. The solid squares and triangles correspond to
observed blue supergiants by Gies and Lambert (\cite{gi92}) and Lennon (\cite{le94}) as
reported by Venn (\cite{ve95b}). The solid circles show the positions of
galactic A--type supergiants observed by Venn (\cite{ve95a}, b). 
}
  \label{NH}
\end{figure}

\subsection{Comparisons with the surface abundances of supergiants}

Most blue supergiants present surface enrichments. For instance,
Walborn (\cite{wa76}, \cite{wa88}) showed that ordinary OB supergiants 
have He-- and N-- enrichments as a result of CNO processing. 
Only the small group of peculiar OBC--stars has the normal 
cosmic abundance ratios (cf. also Howarth and Prinja \cite{ho89}; 
Herrero et al. \cite{He92}; Gies and Lambert \cite{gi92}; Lennon \cite{le94}). 
A possibility 
to explain the He-- and N--enrichments 
in supergiants would be that blue supergiants are on a
blue loop after a first red supergiant stage where
dredge--up has occurred producing the observed surface
enrichments. However as we shall see below, this does not appear
as the good explanantion at least for some of the observed enrichments. 

Fig.~\ref{NH} illustrates the changes of the nitrogen to carbon ratios N/C from the 
ZAMS to the red supergiant stage for stars in the mass range from 9 to 20 M$_{\odot}$.
The N/C ratio appears as the most sensitive
observable parameter. For non--rotating stars, the surface enrichment in nitrogen
only occurs when the star reaches the red supergiant phase;
there, CNO elements are dredged--up by deep convection. The
behaviour is the same as for the He--enrichments discussed above.
For rotating stars, N--excesses already occur during
the MS phase and they are larger for higher rotation and initial
stellar masses. At the end of the MS phase of the 12 M$_\odot$ model, the N/C ratio
is enhanced by factors 2.4 and 4.3
for $v_{\rm ini}$ = 200 and 300 km s$^{-1}$ respectively. These factors are increased
up to 4 and 7.1 for the 20 M$_\odot$ models with the same $v_{\rm ini}$.

On Fig.~\ref{NH} we plot also some observations of supergiants performed by
Gies and Lambert (\cite{gi92}), Lennon (\cite{le94}) and Venn (\cite{ve95a}b). 
The plotted values are $\Delta \log {\rm N/C}=[{\rm N/C}]_*-[{\rm N/C}]_{B}$,
where the abundance ratios in number are measured at the surface of the star ($*$), and at
the surface
of main sequence B--stars ($B$) supposed to have retained
their pristine cosmic abundances. Gies and Lambert (\cite{gi92}) provide ``LTE'' and ``NLTE''
N/C ratios. We plot here the smaller ``LTE'' ratios since, for the stars in common with the Lennon sample,
they are similar to Lennon's results (see the discussion in Venn \cite{ve95b}).
From the positions of the observed stars in the $\log g_{\rm eff}$ versus $\log T_{\rm eff}$ diagram, one obtains that the range of initial masses for the A--type supergiants shown on Fig.~\ref{NH} are
between 5 and 20 M$_\odot$.  

From Fig.~\ref{NH}, one notes first that at the $T_{\rm eff}$
of the observed supergiants, non--rotating stellar 
models predict no enrichment at all unless there is a blue loop.
However it is not likely that all the observed points are at the tip of a blue loop.
Indeed some of the stars have initial masses above 15 M$_\odot$ and, 
at solar metallicity, current grids of models
(Arnett \cite{ar91}; Schaller et al. \cite{sch92}; Alongi et al. \cite{al93};  
Brocato and Castellani \cite{br93}) only predict blue loops for
masses equal or lower than 15 M$_\odot$. Moreover
many of the observed stars have N/C ratios too low to result from a first dredge--up
episode (see for instance the position of the blue loop of the non--rotating 9 M$_\odot$
model in Fig.~\ref{NH}). Therefore Venn (\cite{ve95b}) suggested that at least those stars presenting
the lowest N/C ratios are
on their way from the MS to the red giant branch and have undergone some mixing
in the early stage of their evolution. If such stars are not at all accounted for by standard
evolutionary tracks,
rotating models can naturally reproduce their observed surface abundances as can be seen
on Fig.\ref{NH}. 

Moreover as already noted above, theory predicts larger excesses for higher masses, 
a result in agreement with the suggestion of Takeda and Takeda--Hidai (\cite{ta95}) 
recently confirmed by
McErlean et al. (\cite{mc99}).

\section{Conclusion}

Mass loss by stellar winds
and rotational mixing in the stellar interior are certainly the two 
hydrodynamical phenomena which most deeply affect the evolution of massive stars.
Far from being a small
refinement in the physics of stellar interior, rotation appears as an
essential ingredient of future grids of stellar models.
In particular among the important points wich are not discussed here but
which will be studied in more details in forthcoming papers are
the effects of rotation on the
population of red and blue supergiants at various
metallicities, on the evolutionary scenarios leading to the formation of Wolf--Rayet stars (Maeder 1987;
Fliegner and Langer 1995; Meynet 2000b) and on the stellar yields
(Heger et al. 2000). These questions are important for 
a better understanding of starbursts regions and
of the chemical evolution of galaxies.

\end{document}